\documentclass[12pt]{article}
\usepackage{axodraw}
\def\stone{$\tilde{t}_1$}
\def\sttwo{$\tilde{t}_2$}

\def\stauone{$\tilde{\tau}_1$}
\def\stautwo{$\tilde{\tau}_2$}

\def\snone{$\tilde{\chi}_1^0$}
\def\sntwo{$\tilde{\chi}_2^0$}
\def\snthree{$\tilde{\chi}_3^0$}
\def\snfour{$\tilde{\chi}_4^0$}

\def\scone{$\tilde{\chi}_1^{\pm}$}
\def\sctwo{$\tilde{\chi}_2^{\pm}$}

\def\lsim{\:\raisebox{-0.5ex}{$\stackrel{\textstyle<}{\sim}$}\:}
\def\gsim{\:\raisebox{-0.5ex}{$\stackrel{\textstyle>}{\sim}$}\:}
\def\beq{\begin{equation}}
\def\eeq{\end{equation}}
\def\beqa{\begin{eqnarray}}
\def\eeqa{\end{eqnarray}}

\begin{document}

\renewcommand{\thefootnote}{\fnsymbol{footnote}}

\mbox{ } \\[-1cm]
\mbox{ }\hfill TUM--HEP--490/02\\
\mbox{ }\hfill hep--ph/0211406\\
\mbox{ }\hfill \today\\

\begin{center}
{\Large\bf Detailed analysis of the decay spectrum of a super--heavy
$X$ particle} \\[8mm] 

Cyrille Barbot and Manuel Drees \\[4mm]
{\it Physik Dept., TU M\"{u}nchen, James Franck Str., D--85748
Garching, Germany} \\
\end{center}

\bigskip
\bigskip
\bigskip

\begin{abstract}
\noindent

Decays of superheavy $X$ particles with mass $M_X \sim 10^{12} -
10^{16}$ GeV have been proposed as origin of the observed ultra high
energy cosmic rays (UHECR). We describe in detail the physics
involved in the different steps of the decay of such a particle.  In
particular, we give for the first time the complete set of splitting
functions needed to model a parton shower in the minimal
supersymmetric extension of the Standard Model (MSSM). We present our
results in the form of fragmentation functions of any (s)particle of
the MSSM into any final stable particle (proton, photon, electron,
three types of neutrino, lightest superparticle LSP) at a virtuality
$Q = M_X$, over a scaled energy range $x \equiv 2 E / M_X \in
[10^{-13}, 1]$. Extending the coverage to such small fractional
energies is necessary since the energy region around $10^{18}$ eV and
below could be of considerable interest in testing this kind of model
for generating UHECR. We explicitly demonstrate that our treatment
conserves energy, and discuss the dependence of the final result on
SUSY parameters. We also show that our results are essentially
independent of the necessary extrapolation of the input fragmentation
functions, which are known only for $x \geq 0.1$, towards small $x$.
Finally, we added a new treatment of the color coherence effects at
very small $x$, using the analytic ``MLLA'' solution. Our computer
code will soon be made available.

\end{abstract}
%

\newpage

\section{Introduction}
\label{sec:introduction}
\setcounter{footnote}{1}

In the last few decades ultrahigh energy cosmic rays (UHECR), with
energy above the so--called GZK cut--off \cite{GZK}, have been
observed \cite{expts}. The classical
``bottom--up'' explanation for the acceleration of CR particles
exploits the electromagnetic fields that are likely to be
present in objects like gamma ray bursters \cite{WaxmanGRB}, ``hot
spots'' of radio-galaxies \cite{Biermann} or near super--massive black
holes in dormant quasars \cite{Boldt}. However, it is difficult to
find objects capable of accelerating protons to energies above
$10^{20}$ eV, partly because the product of field strength and spatial
extension of the field does not seem to be sufficiently large, and
partly because the accelerated particles can loose a fair fraction of
their energy in synchrotron radiation. As an attractive alternative,
``top--down'' models have been proposed. Here one postulates the
existence of ``new physics'' at a very high energy scale, i.e. the
existence of super--heavy particles of masses greater than $10^{12}$
GeV, which could decay and hence produce the observed UHECR; this idea
has been reviewed e.g. in \cite{reviewSigl,reviewSarkar}. The
existence of such a very high energy scale strongly suggests the
existence of superparticles with masses not much above 1 TeV in order
to guarantee the perturbative stability of the hierarchy between $M_X$
and the weak scale. We therefore usually allow superparticles as well
as ordinary particles to be produced in $X$ decays, as described by the
minimal supersymmetric extension of the Standard Model (MSSM).

Among the top--down scenarios, we have to distinguish between (at
least) two classes of $X$ particles: GUT particles (gauge or Higgs
bosons of masses close to the GUT scale $\sim 10^{16}$ GeV) could be
trapped in topological defects (cosmic strings or monopoles) formed
during a phase transition in the early Universe, and might be released
when these defects annihilate or collapse \cite{Hill,Monolaces}.
Another possibility is to consider metastable $X$ particles
\cite{wimpzillas} that are distributed freely in the Universe; they
might have been created at the end of inflation \cite{creat}. If their
lifetime $\tau_X$ exceeds the age of the Universe by about a factor
$10^{10} \cdot ( 10^{12} \ {\rm GeV}) / M_X$, the $X$ density required
to explain the observed flux of UHECR events also makes $X$ a good
cold Dark Matter candidate \cite{SHDM, Birkel}. Depending on the model
we are considering, $X$ decay modes can be very different. For
example, GUT gauge bosons will undergo two--body decays into all light
(MS)SM particles with comparable branching ratios. On the other hand,
if $X$ decays only proceed through higher dimensional operators, $X$
will usually undergo many--body decays. In another recently proposed
``brane world'' scenario \cite{Hagiwara}, $\Gamma_X$ is exponentially
suppressed by the relatively large distance between two branes, and
$X$ decays predominantly into light Higgs bosons and their
superpartners. In any case, it is reasonable to assume that $X$ will
decay into some light particles contained in the spectrum of the
MSSM. These primary decay products initiate parton cascades, the
development of which can be predicted from the ``known'' interactions
contained in the MSSM.

Here we describe in detail the calculation of the spectrum of stable
particles (protons, electrons, photons, three kinds of neutrinos, and
lightest superparticles LSP) produced in $X$ decays; a brief summary
of these results has appeared in \cite{BarbotDrees:1}. The first
calculations of this kind \cite{Hill} used simple scaling
fragmentation functions to describe the transition from partons to
hadrons. Later analyses \cite{Birkel,Berezinsky:2000} used Monte Carlo
programs to describe the cascade. However, since we can only expect to
see a single particle from any given cascade, we only need to know the
one--particle inclusive decay spectrum of $X$. This is encoded in
fragmentation functions; the evolution of the cascade corresponds to
the scale dependence of these FFs, which is described by generalized
DGLAP equations \cite{AP}. In contrast to earlier analyses using this
technique, which only include (SUSY) QCD \cite{Rubin:1999,
Coriano:2001, Sarkar:2001, FodorKatz}, or at best a partial treatment
of electroweak interactions \cite{ToldraLSP, Berezinsky:2002}, we
consider all gauge interactions as well as third generation Yukawa
interactions; note that at energies above $10^{20}$ eV, due to the
running of the coupling constants, all gauge interactions are of
comparable strength. We will show that the inclusion of electroweak
gauge interactions in the shower gives rise to a significant flux of
very energetic photons and leptons, beyond the highest proton
energies. Moreover, we carefully model decays of all unstable
particles. As a result, we are for the first time able to fully
account for the energy released in $X$ decay.  We cover all possible
primary $X$ decay modes, i.e. our results should be applicable to all
models where physics at energies below $M_X$ is described by the MSSM.

Before going into technical details, we briefly outline the physics
involved in the decay of an ultra--heavy $X$ particle; it is
summarized in fig.~\ref{X_decay}. By assumption its primary decay is
into 2 or more particles of the MSSM. These primary decay products
will generally not be on--shell; instead, they have very large
(time--like) virtualities, of order $M_X$. Each particle produced in
the primary decay will therefore initiate a parton shower. The basic
mechanism driving the shower development is the splitting of a virtual
particle into two other particles with (much) smaller virtualities;
the dynamics of this process is described by a set of splitting
functions (SFs). As long as the virtuality is larger than the typical
sparticle mass scale $M_{\rm SUSY}$, all MSSM particles participate in
this shower. At virtuality $M_{\rm SUSY} \sim 1$ TeV the breaking of
both supersymmetry and of $SU(2) \times U(1)_Y$ gauge invariance
becomes important. All the massive superparticles that have been
produced at this stage can now be considered to be on--shell, and will
decay into Standard Model (SM) particles and the only (possibly)
stable sparticle, the LSP. The same is true for the heavy SM
particles, i.e. the top quarks and the massive bosons. However, the
lighter quarks and gluons will continue a perturbative parton shower
until they have reached either their on--shell mass scale or the
typical scale of hadronization $Q_{\rm had} \sim 1$ GeV.  At this
stage, strong interactions become non--perturbative, forcing partons
to hadronize into colorless mesons or baryons.  Finally, the unstable
hadrons and leptons will also decay, and only the stable particles
will remain. The spectra of these particles constitute the result of
our calculation, which gives the UHECR spectrum at the location of $X$
decay. Of course, the spectrum on Earth might be modified considerably
due to propagation through the (extra)galactic medium
\cite{reviewSigl}; we will not address this issue here.

Technically the shower development is described through fragmentation
functions (FFs). The dependence of these functions on the virtuality
is governed by the DGLAP evolution equations \cite{AP} extended to
include the complete spectrum of the MSSM. All splitting functions
needed in this calculation are collected in Appendix A.  We
numerically solved the evolution equations for the FFs of any particle
of the MSSM into any other. At scale $M_{\rm SUSY}$ we applied unitary
transformations to the FFs of the unbroken fields (``current
eigenstates'') in order to obtain those of the physical particles
(``mass eigenstates''); details are given in Appendix B. We then model
the decays of all particles and superparticles with mass $\sim M_{\rm
SUSY}$, using the public

\vspace*{5mm}
\begin{center} 
\begin{figure}
\label{X_decay}
\begin{picture}(-400,300)(0,-190)
\SetOffset(500,0)
\SetPFont{Helvetica}{24}
\PText(-405,4)(0)[]{X}
\DashLine(-410,0)(-440,0){3} \Text(-425,7)[]{$\tilde{q}_L$} 
\DashLine(-400,0)(-370,0){3} \Text(-385,7)[]{$\tilde{q}_L$} 
\Line(-370,1)(-340,26)
\Line(-370,0)(-340,25) \Text(-360,25)[]{$\tilde{g}$} 
\Line(-340,26)(-310,46)
\Line(-340,25)(-310,45) \Text(-330,45)[]{$\tilde{g}$}
\Gluon(-340,25)(-310,10){-3}{4}\Text(-320,25)[]{$g$} 
\DashLine(-310,45)(-280,55){3}  \Text(-295,60)[]{$\tilde{q}_L$} 
\GCirc(-280,55){3}{0}
\Line(-310,45)(-280,35) \Text(-295,30)[]{$q_L$} 
\DashLine(-280,-120)(-280,110){4} \Text(-280,-130)[c]{1 TeV}
\Text(-280,-145)[c]{(SUSY}
\Text(-280,-160)[c]{+ $SU(2)\otimes U(1)$}
\Text(-280,-175)[c] {breaking)}

\Line(-280,55)(-250,80) \Text(-265,80)[]{$q$} 
\Line(-250,80)(-190,70) \Text(-220,68)[]{$q$} 
\Gluon(-250,80)(-220,90){3}{4} \Text(-235,97)[]{$g$} 
\Line(-220,90)(-190,100) \Text(-205,103)[]{$q$} 
\Line(-220,90)(-190,80) \Text(-205,80)[]{$q$} 

\Line(-280,56)(-250,46)
\Line(-280,55)(-250,45)
\Text(-265,42)[]{$\tilde{\chi}_2^0$}
\GCirc(-250,45){3}{0}
\Line(-250,45)(-220,55) \Text(-235,57)[]{$q$}
\Line(-250,46)(-170,36)
\Line(-250,45)(-170,35) \Text(-166,35)[l]{$\tilde{\chi}_1^0$}
\Line(-250,45)(-220,25) \Text(-235,27)[]{$q$} 
\Line(-220,55)(-190,65) 
\Gluon(-220,55)(-190,45){-3}{4} 
\DashLine(-190,-120)(-190,110){4} \Text(-190,-130)[c]{1 GeV}
\Text(-190,-145)[c]{(hadronization)}

\Line(-370,0)(-340,-25) \Text(-360,-18)[]{$q_L$} 
\Photon(-340,-25)(-310,-5){3}{5} \Text(-330,-5)[]{$B$} 
\DashLine(-310,-5)(-280,5){3} \Text(-295,8)[]{$\tilde{q}_R$}
\GCirc(-280,5){3}{0}
\DashLine(-310,-5)(-280,-15){3} \Text(-295,-18)[]{$\tilde{q}_R$}
\GCirc(-280,-15){3}{0}
\Line(-340,-25)(-310,-45) \Text(-330,-40)[]{$q$} 
\Photon(-310,-45)(-280,-25){3}{5} \Text(-308,-31)[]{$W$} 
\GCirc(-280,-25){3}{0}
\Line(-280,-25)(-190,-5) \Text(-240,-8)[]{$\tau$} 
\GCirc(-190,-5){3}{0}
\Line(-190,-5)(-160,5) \Text(-175,9)[]{$a_1^{-}$}
\GCirc(-160,5){3}{0}
\Line(-160,5)(-130,15) \Text(-145,17)[]{$\rho^{-}$}
\GCirc(-130,15){3}{0}
\Line(-130,15)(-100,15) \Text(-115,11)[]{$\pi^{-}$}
\GCirc(-100,15){3}{0}
\Line(-100,15)(-70,15) \Text(-66,15)[l]{$\nu_\mu$} 
\Line(-100,15)(-70,5) \Text(-78,2)[]{$\mu^{-}$} 
\GCirc(-70,5){3}{0}
\Line(-70,5)(-50,5) \Text(-46,5)[l]{$\nu_\mu$} 
\Line(-70,5)(-50,-5) \Text(-46,-7)[l]{$\nu_e$} 
\Line(-70,5)(-50,-15) \Text(-46,-16)[l]{$e^{-}$} 

\Line(-130,15)(-100,35) \Text(-115,35)[]{$\pi^0$} 
\GCirc(-100,35){3}{0}
\Photon(-100,35)(-70,35){3}{5} \Text(-66,35)[]{$\gamma$} 
\Photon(-100,35)(-70,55){3}{5} \Text(-66,57)[]{$\gamma$} 

\Line(-160,5)(-130,-5) \Text(-145,-8)[]{$\pi^0$}
\GCirc(-130,-5){3}{0}
\Photon(-130,-5)(-100,-25){3}{5} \Text(-96,-27)[]{$\gamma$} 
\Photon(-130,-5)(-100,-5){3}{5} \Text(-96,-3)[]{$\gamma$} 
\Line(-190,-5)(-160,-15) \Text(-175,-16)[]{$\nu_\tau$} 
\Line(-280,-25)(-100,-45) \Text(-96,-45)[l]{$\nu_\tau$} 

\Line(-310,-45)(-280,-65) \Text(-300,-62)[]{$q$} 
\Line(-280,-65)(-250,-55) \Text(-265,-53)[]{$q$} 
\Gluon(-280,-65)(-250,-75){-3}{4} \Text(-265,-80)[]{$g$}
\Gluon(-250,-75)(-220,-95){-3}{4} \Text(-240,-95)[]{$g$} 
\Line(-220,-95)(-190,-105) \Text(-205,-105)[]{$q$} 
\Line(-220,-95)(-190,-85) \Text(-205,-85)[]{$q$}

\Gluon(-250,-75)(-220,-65){3}{4} \Text(-240,-62)[]{$g$} 
\Line(-220,-65)(-190,-55)  \Text(-205,-55)[]{$q$}
\Line(-220,-65)(-190,-75)  \Text(-205,-75)[]{$q$}

\GOval(-190,65)(20,7)(0){0} 
\GCirc(-170,65){3}{0}
\Text(-175,65)[r]{$n$}
\Line(-170,65)(-130,80) \Text(-126,82)[l]{$p$}
\Line(-170,65)(-130,65) \Text(-126,65)[l]{$e^{-}$}
\Line(-170,65)(-130,50) \Text(-126,48)[l]{$\nu_e$} 

\GOval(-190,-80)(15,7)(0){0} 
\GCirc(-168,-83){3}{0}
\Text(-170,-80)[r]{$\pi^0$}
\Photon(-168,-83)(-130,-65){3}{5} \Text(-126,-63)[]{$\gamma$} 
\Photon(-168,-83)(-130,-95){3}{5} \Text(-126,-98)[]{$\gamma$} 

\end{picture} 
\caption{Schematic MSSM cascade for an initial squark with a
virtuality $Q \simeq M_X$. The full circles indicate decays of massive
particles, in distinction to fragmentation vertices. The two vertical
dashed lines separate different epochs of the evolution of the
cascade: at virtuality $Q > M_{\rm SUSY}$, all MSSM particles can be
produced in fragmentation processes. Particles with mass of order
$M_{\rm SUSY}$ decay at the first vertical line. For $M_{\rm SUSY} > Q
> Q_{\rm had}$ light QCD degrees of freedom still contribute to the
perturbative evolution of the cascade. At the second vertical line,
all partons hadronize, and unstable hadrons and leptons decay. See the
text for further details.}
\end{figure}
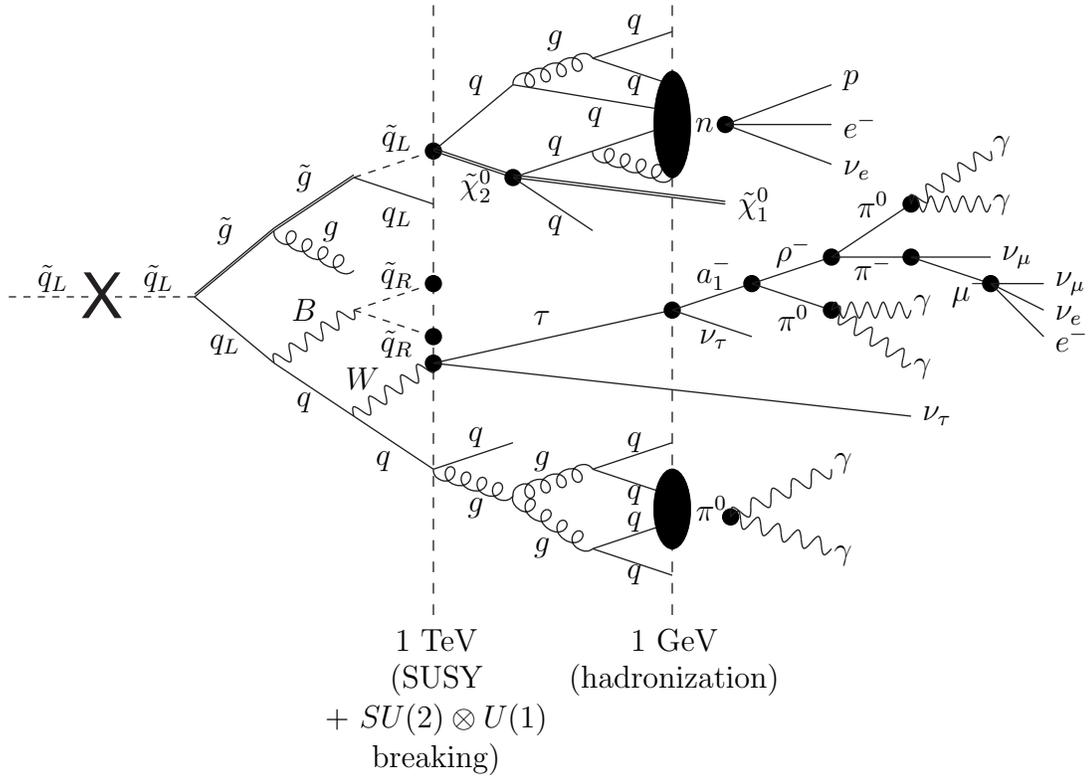
\end{center}
\vspace*{5mm}

\noindent
code ISASUSY \cite{Isasusy} to compute the branching ratios for all
allowed decays, for a given set of SUSY parameters. If R-parity is
conserved, we obtain the final spectrum of the stable LSP at this
step; the rest of the available energy is distributed between the SM
particles. After a second perturbative cascade down to virtuality
$\sim \max(m_q,Q_{\rm had})$, the quarks and gluons will hadronize, as
stated before. This non--perturbative phenomenon is parameterized in
terms of ``input'' FFs. We use the results of ref.\cite{Poetter},
which are based on fits to LEP data. We paid special attention to the
conservation of energy; this was not possible in previous studies,
because of the incomplete treatment of the decays of particles with
mass of order $M_{\rm SUSY}$. We are able to check energy conservation
at each step of the calculation, up to a numerical accuracy of a few
per mille.

The remainder of this article is organized as follows. In sec.~2 we
describe the technical aspects of the calculation. The derivation and
solution of the evolution equations is outlined. We also check that
our final results are not sensitive to the necessary extrapolation of
the input FFs. Numerical results are presented in sec.~3. We give the
energy fractions carried by the seven stable particles for any primary
$X$ decay product, and study the dependence of our results on the SUSY
parameters. We finally describe our implementation of color coherence
effects at small $x$ using the modified leading log approximation (MLLA).
Sec.~4 is devoted to a brief summary and conclusions. Technical
details are delegated to a series of Appendices, giving the complete
list of splitting functions, the unitary transformations from the
interaction states to the physical states, our treatment of 2-- and
3--body decays, parameterizations of the input FFs, and finally a
complete set of FFs obtained with our program for a given set of SUSY
parameters (corresponding to a gaugino--like LSP with a low value of
$\tan \beta \sim 3.6$ and $M_{\rm SUSY} \sim 500$ GeV).

\section{Technical aspects of the calculation}
\label{sec:technical}

In this section we describe how to calculate the spectra of stable
particles produced in $X$ decays: protons, electrons, photons, the
three types of neutrinos and LSPs, and their antiparticles. Note that
at most one out of the many particles produced in a typical $X$ decay
will be observed on Earth. This means that we cannot possibly measure
any correlation between different particles in the shower; the energy
spectra of the final stable particles are indeed the only measurable
quantities. These spectra are given by the differential decay rates $d
\Gamma_X / d E_P$, where $P$ labels the stable particle we are
interested in. This is a well--known problem in QCD, where parton
showers were first studied. The resulting spectrum can be written in
the form \cite{QCDrev}
\beq \label{def_ff}
\frac {d \Gamma_X} {d x_P} = \sum_I \frac {d \Gamma(X \rightarrow I)}
{d x_I} \otimes D^P_I(\frac{x_P}{x_I}, M_X^2),
\eeq
where $I$ labels the MSSM particles into which $X$ can decay, and we
have introduced the scaled energy variable $x = 2 E / M_X$. $d
\Gamma(X \rightarrow I) / d x_I$ depends on the phase space in a
particular decay mode; for a two--body decay, $d \Gamma(X \rightarrow
I) / d x_I \propto \delta(1-x_I)$. The convolution is defined as
\beq \label{conv}
f(z) \otimes g(x/z) = \int_{x}^{1} f(z) g \left( \frac {x} {z} \right)
\, \frac{dz}{z}.
\eeq
All the nontrivial physics is now contained in the fragmentation
functions (FFs) $D^P_I(z,Q^2)$. They encode the probability for a
particle $P$ to originate from the shower initiated by another
particle $I$, where the latter has been produced with initial
virtuality $Q$. This implies the ``boundary conditions''
\beq \label{boundary}
D^J_I(z,m_J^2) = \delta_I^J \cdot \delta(1-z),
\eeq
which simply say that an on--shell particle cannot participate in the
shower any more. As already explained in the Introduction, for $Q >
M_{\rm SUSY}$ all MSSM particles $J$ are active in the shower, and
thus have to be included in the list of ``fragmentation products''.

The evolution of the FFs with increasing virtuality is described by the
well--known DGLAP equations \cite{AP}. In the next two subsections
we discuss these evolution equations, and their solution, in more
detail. We first only include strong (SUSY--QCD) interactions.
However, at energies above $10^{20}$ eV all gauge interactions are of
comparable strength. The same is true for interactions due to the
Yukawa coupling of the top quark, and possibly also for those of the
bottom quark and tau lepton. In a second step we therefore extend the
evolution equations to include these six different interactions. We
then describe the decays of heavy (s)particles, which happen at
virtuality $Q = M_{\rm SUSY} \sim 1$ TeV. At $Q < M_{\rm SUSY}$ only
QCD interactions need to be included, greatly simplifying the
treatment of the evolution equations in this domain. Finally, we
describe the nonperturbative hadronization, and the weak decays of
unstable hadrons and leptons.

\subsection{Evolution equations in QCD and SUSY--QCD}

For convenience, we review here the DGLAP evolution equations in
ordinary QCD. As already noted, the FF $D_{p}^P (x,Q^2)$ of a parton
(quark or gluon) $p$ into a particle (parton or hadron) $P$ describes
the probability of fragmentation of $p$ into $P$ carrying energy $E_P
= xE_p$ at a virtuality scale $Q$. If $P$ is itself a parton, the FF
has to obey the boundary condition (\ref{boundary}). However, if $P$
is a hadron, the $x-$dependence of the FF cannot be computed
perturbatively; it is usually derived from fits to experimental
data. Perturbation theory does predict the dependence of the FFs on
the virtuality $Q$: it is described by a set of coupled
integro--differential equations. In leading order (LO), these QCD
DGLAP evolution equations can be written as \cite{QCDrev} :
\beqa \label{e1}
\frac {dD_{q_i}^P (x,Q^2)} {d\log(Q^2)}  &=&
\frac {\alpha_S(Q^2)} {2\pi} \left\{ P_{gq}(z) \otimes
  D_{g}^{P}(\frac{x}{z},Q^2) + P_{qq}(z) \otimes D_{q_i}^P
(\frac{x}{z},Q^2) \right\}\,, \nonumber \\
\frac {dD_g^P(x,Q^2)} {d\log(Q^2)}
&=& \frac {\alpha_{S}(Q^2)} {2\pi} \left\{ P_{gg}(z) \otimes
  D_{g}^P(\frac{x}{z},Q^2) + \sum_{i=1}^{2F} P_{qg}(x) \otimes
D_{q_i}^P (\frac{x}{z}, Q^2) \right\}\,, \eeqa
where $\alpha_S$ is the running QCD coupling constant, $F$ is the
number of active flavors (i.e. the number of Dirac quarks whose mass
is lower than $Q$), and $i$ labels the quarks and
antiquarks.\footnote{Note that the DGLAP equations given here are the
{\it time--like} ones, which describe the evolution of fragmentation
functions. In leading order they differ from the space--like DGLAP
equation (describing the evolution of distribution functions of
partons inside hadrons) only through a transposition of the matrix of
the splitting functions.} The convolution has been defined in
eq.(\ref{conv}). The physical content of these equations can be
understood as follows. A virtual quark $q_i$ can reduce its
virtuality by emitting a gluon; the final state then contains a quark
and a gluon. Either of these partons (with reduced virtuality) can
fragment into the desired particle $P$; this explains the occurrence of
two terms in the first eq.(\ref{e1}). Analogously, a gluon can either
split into two gluons, or into a quark--antiquark pair, giving rise to
the two terms in the second eq.(\ref{e1}).

These partonic branching processes are described by the splitting
functions (SFs) $P_{p_2 p_1}(x)$, for parton $p_1$ splitting into
parton $p_2$, where $x = E_{p_2} / E_{p_1}$. As already noted, in pure
QCD there are only three such processes: gluon emission off a quark or
gluon, and gluon splitting into a $q \bar q$ pair. The first of these
processes gives rise to both SFs appearing in the first eq.(\ref{e1});
momentum conservation then implies $P_{qq}(x) = P_{gq}(1-x)$, for $x
\neq 1$. Similarly, $P_{gg}(x) = P_{gg}(1-x)$ and $P_{qg}(x) =
P_{qg}(1-x)$ follows from the symmetry of the final states resulting
from the splitting of a gluon. Special care must be taken as $x
\rightarrow 1$. Here one encounters infrared singularities, which
cancel against virtual quantum corrections. The physical result of
this cancellation is that the energy of the fragmenting parton $p$ is
conserved, which requires 
\beq \label{momcons}
\sum_P \int_0^1 x D_p^P(x,Q^2) = 1 \,\,\, \forall p, Q^2.
\eeq
This can be ensured, if
\beq \label{sumrule}
\int_0^1 dx\, x \sum_{p'} P_{p'p}(x) = 0 \,\,\, \forall p.
\eeq
Note that these integrals must give zero (rather than one), since
eqs.(\ref{e1}) only describe the {\em change} of the FFs. The explicit
form of the QCD SFs is \cite{AP}:
\beqa
\label{e3}
P_{qq}(x) &=& \frac{4}{3} \left( \frac {1+x^2} {1-x} \right )_+\,,
\nonumber \\
P_{gq}(x) &=& \frac{4}{3} \frac {1 + (1-x)^2} {x}\,, 
\nonumber \\
P_{qg}(x) &=& \frac{1}{2} \left[ (1-x)^2 + x^2 \right] \,,
\nonumber \\
P_{gg}(x) &=& 6 \left[ \frac {1-x} {x} + x(1-x) + \frac {x} {(1-x)_+}
+ \delta(1-x) \left( \frac{11}{12} - \frac{F}{18} \right) \right] \,.
\eeqa
The ``+'' distribution, which results from the cancellation of $x
\rightarrow 1$ divergences as outlined above, is defined as:
\beq \label{e4} 
\int_{0}^{1} f(x) g(x)_+ dx = \int_{0}^{1} [f(x) - f(1)] g(x),
\eeq
while $g(x)_+ = g(x)$ for $x \neq 1$. Finally, the scale dependence of
$\alpha_S(Q^2)$ is described by the following solution of the relevant
renormalization group equation (RGE):
\beq \label{alphas}
\alpha_S(Q^2) = \frac{2\pi B}{\log\frac{Q^2}{\Lambda^2}} \,,
\eeq
where $\Lambda \sim 200$ MeV is the QCD scale parameter, and $B =
6/(33 - 2F)$. 

Note that eqs.(\ref{e1}) list different FFs for all (anti)quark flavors
$q_i$. At first sight it thus seems that one has to deal with a system
of $2F+1$ coupled equations. In practice the situation can be
simplified considerably by using the linearity of the evolution
equations. This implies 
\beq \label{split}
D_p^P(x,Q^2) = \sum_{p'} \tilde D_p^{p'}(z, Q^2, Q_0^2) \otimes
D_{p'}^P(\frac{x}{z}, Q_0^2),
\eeq
where the generalized FFs $\tilde D_p^{p'}$ again obey the evolution
equations (\ref{e1}). Moreover, they satisfy the boundary conditions
$D_p^{p'}(x,Q_0^2, Q_0^2) = \delta(1-x) \delta_p^{p'}$ at some
convenient value of $Q_0 < Q$. The $\tilde D$ thus describe the purely
perturbative evolution of the shower between virtualities $Q$ and
$Q_0$. This ansatz simplifies our task, since all quark flavors have
exactly the same strong interactions, i.e. we can use the {\em same}
$\tilde D_{q_i}^{p'}$ for all quarks $q_i$ with $m_{q_i} < Q_0$.
Moreover, we only have to distinguish three different cases for $p': \
q_i, q_j$ with $j \neq i$, and $g$. All flavor dependence is then
described by the $D_{p'}^P(x, Q_0^2)$; for sufficiently small $Q_0$,
these can be taken directly from fits to experimental data. If we make
the additional simplifying assumption that all quarks and antiquarks
are produced with equal probability in primary $X$ decays, we
effectively only have to introduce two generalized FFs $\tilde D$ for
a given particle $P$, one for the fragmentation of gluons and one for
the fragmentation of any quark. In other words, in pure QCD we only
need to solve a system of two coupled equations.

The introduction of squarks $\tilde q_i$ and gluinos $\tilde g$,
i.e. the extension to SUSY--QCD, requires the introduction of FFs
$D_{\tilde q_i}^P, \ D_{\tilde g}^P$. This gives rise to new SFs,
describing the emission of a gluon by a squark or gluino, as well as
splittings of the type $q_i \rightarrow \tilde q_i \tilde g, \ \tilde
q_i \rightarrow q_i \tilde g$ and $\tilde g \rightarrow \tilde q_i
\bar q_i$. We thus see that any of the four types of partons $(q_i,
\tilde q_i, g, \tilde g)$ can split into any (other) parton. The
complete set of evolution equations thus contains 16 SFs \cite{Jones},
which we collect in Appendix A. The presence of new particles with
$SU(3)$ interactions also modifies the running of $\alpha_S$. One can
still use eq.(\ref{alphas}), but now $B_{SUSY} = 2/(9 - F)$.

\subsection{Evolution equations in the MSSM}

We now extend our discussion of the evolution equations to the full
MSSM. We already saw in the Introduction that superparticles can only
be active in the shower evolution at virtualities $Q > M_{\rm SUSY}
\sim 1$ TeV. This means that the supersymmetric part of the shower
evolution can be described in terms of generalized FFs $\tilde D_I^J$
satisfying the boundary condition 
\beq \label{boundary_MSSM}
\tilde D_I^J(x,M^2_{\rm SUSY},M^2_{\rm SUSY}) = \delta_I^J \delta(1-x),
\eeq
where $I$ and $J$ label any (s)particle contained in the MSSM. Note
that eq.(\ref{boundary_MSSM}) differs from eq.(\ref{boundary}) since
the former is valid for {\em all} particles in the MSSM, including
light partons. According to the discussion following eq.(\ref{split})
we only have to consider those particles to be distinct that have
different interactions. We include all gauge interactions in this part
of the shower evolution, as well as the Yukawa interactions of third
generation (s)fermions and Higgs bosons, but we ignore first and
second generation Yukawa couplings, as well as all interactions
between different generations. This immediately implies that we do not
need to distinguish between first and second generation particles.
Moreover, we ignore CP violation, which means that we need not
distinguish between particles and antiparticles. Finally, the
electroweak $SU(2)$ symmetry can be taken to be exact at virtuality $Q
> M_{\rm SUSY}$, i.e. we need not distinguish between members of the
same $SU(2)$ multiplet. This is analogous to ordinary QCD, where one
does not need to introduce different FFs for quarks with different
colors. Our assumption implies that $X$ is an $SU(2)$ singlet. Had we
allowed \cite{Berezinsky:2002} $X$ to transform nontrivially under
$SU(2)$, the $SU(2)$ splitting functions would have to be modified
\cite{CCC}.\footnote{A minor caveat might be in order here. We assume
the electroweak part of the evolution of the FFs to be described by
the equations for an unbroken $SU(2) \times U(1)_Y$ theory. It seems
physically reasonable that this should be a good approximation as long
as both the energy and the virtuality of the gauge bosons is large
compared to their masses, which is the case in our
calculation. Indeed, existing loop calculations \cite{newloop} confirm
that the leading logarithmic corrections, which we treat here, are the
same in the broken and unbroken theory, as long as all scales are
large compared to $M_W$. However, these calculations all deal with the
electroweak equivalent of structure functions, rather than
fragmentation functions. Strictly speaking our assumption, although
reasonable, has therefore not yet been backed up by an explicit
calculation that includes massive gauge bosons.}

Altogether we therefore need to treat 30 distinct particles: six
quarks $q_L, u_R, d_R, t_L, t_R, b_R$, four leptons $l_L, e_R, \tau_L,
\tau_R$, three gauge bosons $B, W, g$, two Higgs bosons $H_1, H_2$,
and all their superpartners; $H_1$ couples to down--type quarks and
leptons, while $H_2$ couples to up--type quarks.  Note that a
``particle'' often really describes the contribution of several
particles which are indistinguishable by our criteria. For example,
the ``quark'' $u_R$ stands for all charge$-2/3$ right--handed quarks
and antiquarks of the two first generations, i.e.  $u_R, \, c_R$ and
their antiparticles $\overline{u}_R, \, \overline{c}_R$. This can be
expressed formally as $D_{u_R}^P = \left( D_{u_R}^P + D_{c_R}^P +
D_{\overline{u}_R}^P + D_{\overline{c}_R}^P \right) / 4$, where in our
approximation the four terms in the sum are all identical to each
other after the final state $P$ has been summed over particle and
antiparticle.\footnote{A consistent interpretation of, e.g., $u_R$ as
a ``particle'' requires that $u_R$ stands for the {\em average} of
$u_R, \, c_R$ etc. when $u_R$ appears as {\em lower} index of a
generalized FF, as described in the text. However, $u_R$ stands for
the {\em sum} of $u_R, \, c_R$ etc. when $u_R$ is an {\em upper} index
of a $\tilde D$. With this definition, we have $\tilde D_{u_R}^{u_R}
(x,M^2_{\rm SUSY}) = \delta(1-x)$. This interpretation also fixes
certain multiplicity factors in the DGLAP equations, as detailed in
Appendix A. This treatment is only possible if $X$ has equal branching
ratio into $u_R, \, c_R$ etc. However, we expect the differences
between decays into first or second generation quarks to be very small
even in models where these branching ratios are not the same.}
Similarly, $q_L$ stands as initial particle for an average over the
two $SU(2)$ quark doublets of the two first generations $(u_L,d_L)$
and $(c_L,s_L)$, and their antiparticles. Note that all group indices
of the particle in question are summed over. In the usual case of QCD
this only includes summation over color indices, but in our case it
includes summation over $SU(2)$ indices, since $SU(2)$ is
(effectively) conserved at energies above $M_{\rm SUSY}$.

Let us first discuss the scale dependence of the six coupling
constants that can affect the shower evolution significantly at scales
$Q > M_{\rm SUSY}$. These are the three gauge couplings $g_Y$, $g_2$
and $g_S$, which are related to the corresponding ``fine structure
constants'' through $\alpha_i \equiv g_i^2 / (4\pi), \ i \in \left\{
Y, 2, S \right\}$. Moreover, the third generation Yukawa couplings are
proportional to the masses of third generation quarks or leptons:
\beqa \label{e6}
y_t &=& \frac{g\,m_t}{\sqrt{2}\,m_W\sin{\beta}}\,,
\nonumber \\
y_b &=& \frac{g\,m_b}{\sqrt{2}\,m_W\cos{\beta}}\,,
\nonumber \\
y_\tau &=& \frac{g\,m_\tau}{\sqrt{2}\,m_W\cos{\beta}}\,,
\eeqa
where $\tan\beta \equiv \langle H_2^0 \rangle / \langle H_1^0
\rangle$. The couplings $y_b$ and $y_\tau$ are only significant if
$\tan\beta \gg 1$. Note that in many models, values $\tan\beta \simeq
m_t(m_t) / m_b(m_t) \simeq 60$ are possible, in which case $y_b$ and
$y_\tau$ are comparable in magnitude to $g_S$ and $g_2$, respectively.
The LO RGEs for these six MSSM couplings are \cite{susyrge}:
\beqa \label{e7}
\frac{dg_Y}{dt} &=& 11\,\frac{g_Y^3}{16\pi^2}\,,
\nonumber \\
\frac{dg_2}{dt} &=& \frac{g_2^3}{16\pi^2}\,,
\nonumber \\
\frac{dg_S}{dt} &=& -3\,\frac{g_S^3}{16\pi^2}\,,
\nonumber \\
\frac{dy_t}{dt} &=& \frac {y_t} {16\pi^2} \left( 6y_t^2 + y_b^2 -
\frac {13} {9} g_Y^2 -3 g_2^2 -\frac {16} {3} g_S^2 \right) \,,
\nonumber \\
\frac{dy_b}{dt} &=& \frac {y_b} {16\pi^2} \left( 6y_b^2 + y_t^2 +
y_\tau^2 - \frac{7}{9} g_Y^2 - 3g_2^2 - \frac{16}{3} g_S^2 \right)\,,
\nonumber \\
\frac{dy_\tau}{dt} &=& \frac {y_\tau} {16\pi^2} \left( 3y_b^2 +
4y_\tau^2 - 3g_Y^2 - 3g_2^2\right)\,,
\eeqa
where $t = \log \frac{Q}{Q_0}$ parameterizes the logarithm of the
virtuality, and $Q_0$ is an arbitrary scale where the numerical values
of these couplings constants are ``known'' (in case of the Yukawa
couplings, up to the dependence on $\tan\beta$). As well known
\cite{gutification}, given their values measured at $Q_0 \simeq 100$
GeV eqs.(\ref{e7}) predict the three gauge couplings to unify at scale
$M_{\rm GUT} \simeq 2 \cdot 10^{16}$ GeV, i.e. $g^2_S(M_{\rm GUT}) =
g^2_2(M_{\rm GUT}) = 5 g^2_Y(M_{\rm GUT}) / 3 \simeq 0.52$, where
the Clebsch--Gordon factor of 5/3 is predicted by most simple unified
groups, e.g. $SU(5)$ or $SO(10)$. We solved these equations by the
Runge--Kutta method; of course, the RGEs for the gauge couplings can
trivially be solved analytically, but the additional numerical effort
required by including eqs.(\ref{e7}) in the set of coupled
differential equations that need to be solved numerically is
negligible.

\setcounter{footnote}{0}
The main numerical effort lies in the solution of the system of 30
coupled DGLAP equations, which are of the form:
\beq
\label{e8}
\frac {d \tilde D_I^J} {d\log(Q^2)} (x,Q^2,M^2_{\rm SUSY}) = \sum_K \frac {\alpha_{KI}
(Q^2)} {2\pi} P_{KI}(z) \otimes \tilde D_K^J(\frac{x}{z} ,Q^2,M^2_{\rm SUSY})\,,
\eeq
where $I,J,K$ run over all the 30 particles, and $\alpha_{KI}(Q^2) =
g_{KI}^2 / 4\pi$ is the (running) coupling constant associated
with the corresponding vertex; note that at this stage we are using
interaction (or current) eigenstates to describe the spectrum.
Generically denoting particles with spin 1, 1/2 and 0 as $V, \ F$ and
$S$ (for vector, fermion and scalar), we have to consider\footnote{We
do not need to consider $S \rightarrow SS$, since the corresponding
dimensionful coupling is ${\cal O}(M_{\rm SUSY}) \ll Q$ in this
domain, i.e. these processes are much slower than the relevant time
scale $1/Q$.} branching processes of the kind $V \rightarrow VV, \ V
\rightarrow FF, \ V \rightarrow SS, \ F \rightarrow FV, \ F
\rightarrow FS, \ S \rightarrow SV$ and $S \rightarrow F F$. All these
branching processes already occur in SUSY--QCD. The splitting
functions can thus essentially be read off from the results of
ref.\cite{Jones}, after correcting for different group [color and/or
$SU(2)$] and multiplicity factors. The coefficients of the
$\delta(1-x)$ terms in diagonal SFs can be fixed using the momentum
conservation constraint in the form (\ref{sumrule}); note that these
constraints have to be satisfied for each of the six interactions
separately. The explicit form of the complete set of MSSM SFs
$P_{KI}(x)$ is given in Appendix A.

We solved these equations numerically using the Runge--Kutta
method. To that end the FFs were represented as cubic splines, using
50 points which were distributed equally on a logarithmic scale in $x$
for $10^{-7} \leq x \leq 0.5$, and 50 additional points distributed
equally in $\log(1-x)$ for $0.5 \leq x \leq 1-10^{-7}$.  Starting from
the boundary conditions\footnote{Technically, these $\delta-$functions
are represented by narrow Gaussians centered at $x=1$, normalized to
give unity after integration over $x \leq 1$.}  (\ref{boundary_MSSM}),
we arrive at the $30\times 30$ generalized fragmentation functions at
virtuality $Q=M_X$. Here we assume that the evolution equations
describe the perturbative cascade at these energies correctly. We will
comment on the limitations of our treatment at the end of this
Section.

\subsection{Evolution of the cascade below $Q = 1$ TeV}
\label{subsec:non_pert}
\setcounter{footnote}{0}

Here we would like to describe the physics at scales at and below
$M_{\rm SUSY}$: the breaking of both supersymmetry and $SU(2)\otimes
U(1)$ symmetry, the decay of unstable (s)particles with masses of
order $M_{\rm SUSY}$, the pure QCD shower evolution down to $Q_{\rm
had}$, the non--perturbative hadronization of quarks and gluons, and
finally the weak decays of unstable leptons and hadrons. For
simplicity we assume that all superparticles, the top quark as well as
the $W, \ Z$ and Higgs bosons all decouple from the shower and decay
at the same scale $M_{\rm SUSY} \simeq 1$ TeV. The fragmentation of
$b$ and $c$ quarks is treated using the boundary condition
(\ref{boundary}) at their respective mass scales of 5 and 1.5 GeV,
while the nonperturbative hadronization of all other partons takes
place at $Q_{\rm had} = 1$ GeV.

At $Q = M_{\rm SUSY}$ we break both Supersymmetry and $SU(2) \otimes
U(1)$.  All (s)particles acquire their masses in this process, and in
many cases mix to give the mass eigenstates. This means that we have
to switch from a description of the particle spectrum in terms of
current eigenstates to a description in terms of physical mass
eigenstates. This is accomplished by unitary transformations of the
type\footnote{Note that the squares of the coefficients $c_{SJ}$
appear in eq.(\ref{trafo}), since the FFs describe probabilities,
which are related to the square of the wave functions of the particles
in question.}
\beq \label{trafo}
\tilde D^S_I = \sum_J |c_{SJ}|^2 \tilde D^J_I\,.
\eeq
Unitarity requires $\sum_S |c_{SJ}|^2 = \sum_J |c_{SJ}|^2 = 1$, if the
current state $J$ has the same number of degrees of freedom as the
physical state $S$.  This is often not the case in the usual
convention; then some care has to be taken in writing down the
$|c_{SJ}|^2$, see Appendix B. We use the following physical particles:
$u\,, d\,, b\,, s\,, c\,, t$ quarks and $e\,, \mu\,, \tau$ leptons now
have both left-- and right--handed components, i.e. they have twice as
many degrees of freedom as the corresponding states with fixed
chirality.  The neutrinos remain unchanged, since we ignore the
interactions of right--handed neutrinos. The gluons also remain
unchanged, since $SU(3)$ remains exact below $M_{\rm SUSY}$. The
electroweak gauge sector of the SM is described by $W := W^+ + W^-$, Z
and $\gamma$; note that the massive gauge bosons absorb the Goldstone
modes of the Higgs sector, and hence receive corresponding
contributions in eq.(\ref{trafo}). The Higgs sector consists of two
charged Higgs bosons $H^{\pm}$ (described by $H = H^+ + H^-$) and the
three neutral ones $H^{0}$, $h^{0}$ and $A^{0}$; the neutral Higgs
bosons are described by real fields, which contain a single degree of
freedom. In the SUSY part of the spectrum, the gluino $\tilde g$ as
well as the first and second generation sfermions $\tilde{u}_{L,R}$,
$\tilde{d}_{L,R}$, $\tilde{s}_{L,R}$, $\tilde{c}_{L,R}$ and sneutrinos
remain unchanged (but $\tilde u_L$ and $\tilde d_L$, etc., are now
distinguishable). The $SU(2)$ singlets and doublets of third
generation charged sfermions mix to form mass eigenstates \stone,
\sttwo, $\tilde b_1, \ \tilde{b}_2$, \stauone, \stautwo. Similarly,
the two Dirac charginos \scone\, and \sctwo are mixtures of charged
higgsinos and winos, and the four Majorana neutralinos \snone, \sntwo,
\snthree, \snfour, in order of increasing masses, are mixtures of
neutral higgsinos, winos and binos.

The numerical values of many of the $c_{SJ}$ depend on the parameters
describing the breaking of supersymmetry. We choose four different
sets of parameters, which describe typical regions of the parameter
space, in order to study the impact of the details of SUSY breaking on
the final spectra. We take two fairly extreme values of $\tan(\beta) =
3.6$ and $48$, and two sets of dimensionful parameters corresponding
to higgsino--like and gaugino--like states \scone, \snone\ and \sntwo.
We used the software ISASUSY \cite{Isasusy} to compute the mass
spectrum and the mixing angles of the sparticles and Higgses for a
given set of SUSY parameters.

Having computed the spectrum of physical (massive) particles, we have
to treat the decay of all unstable particles with mass near $M_{\rm
SUSY}$. Since we assumed $R-$parity to be conserved, the lightest
supersymmetric particle (LSP) is stable. In our four scenarios (as in
most of parameter space) the LSP is the lightest neutralino \snone.
The end products of these decays are thus light SM particles and
LSPs. Note that decays of heavy sparticles often proceed via a
cascade, where the LSP is produced only in the second, third or even
fourth step, e.g. $\tilde g \rightarrow \bar u \tilde u_L \rightarrow
\bar u d \tilde \chi_1^+ \rightarrow \bar u d e^+ \nu_e \tilde
\chi_1^0$. In order to model these decays we again use ISASUSY, which
computes the branching ratios for all allowed tree--level 2-- and
3--body decay modes of the unstable sparticles, of the top quark and
of the Higgs bosons. Together with the known branching ratios of the
$W$ and $Z$ bosons, this allows us to compute the spectra of the SM
particles and the LSP after all decays, by convoluting the spectra of
the decaying particles with the energy distributions calculated for
2-- or 3--body decays. The total generalized FF of any MSSM current
eigenstate $I$ into a light or stable physical particle $s$ (quark,
gluon, lepton, photon or LSP) is then
\beq \label{decay}
\tilde D_I^s = \tilde D_I^{S=s} + \sum_{S \neq s} \tilde D_I^S \otimes
\tilde P_{sS},
\eeq
where $\tilde P_{sS}$ describes the spectrum of $s$ in the decay $S
\rightarrow s$. We compute these spectra from phase space, including
all mass effects, but we didn't include the matrix elements. The
spectra for each decay mode of the heavy particle $S$ are normalized
to give the correct branching ratio, as computed by ISAJET. As far as
LSPs are concerned, eq.(\ref{decay}) already gives the final result,
i.e. $D_I^{\rm LSP} = \tilde D_I^{\rm LSP}$. If $s$ is a lepton or
photon, eq.(\ref{decay}) describes the FF at all virtualities between
$M_{\rm SUSY}$ and $m_b = 5$ GeV.

As we will see shortly, in some cases two--body decays can lead to
sharp edges in the FFs at intermediate values of $x$. This can happen
if the primary decay product is a massive particle with only weak
interactions. In that case a substantial fraction of the initial
$\delta-$peak at $x=1$ survives even after the evolution; convolution
of this $\delta-$peak with a two--body decay distribution leads to a
flat $x$ distribution of the decay products between some $x_{\rm min}$
and $x_{\rm max}$. An accurate description of these contributions to
the FFs sometimes requires the introduction of additional points near
$x_{\rm min}$ and/or $x_{\rm max}$ in the splines describing these
FFs.

The perturbative evolution in the QCD sector does not stop at $M_{\rm
SUSY}$, but continues until virtuality $Q_0 = \max(m_q, Q_{\rm
had})$. This part can be treated by introducing generalized FFs
$\tilde D_p^{p'}$ as in eq.(\ref{split}), where $(p, p') \in
\{u,d,s,c,b,g\}$ are light QCD partons. We use once more the DGLAP
evolution equations, but this time for pure QCD, evolving these
generalized FFs between $Q_0$ and $M_{\rm SUSY}$. The generalized
partonic FFs between $Q_0$ and $M_X$ can then be computed through one
more convolution:
\beq \label{dofq0}
\tilde D_I^p(x,M_X^2,Q_0^2) = \sum_{p'} \tilde D_I^{p'}(z,M_X^2,M_{SUSY}^2)
\otimes \tilde D_{p'}^p(\frac{x}{z},M_{SUSY}^2,Q_0^2) \,.
\eeq

The total partonic FFs at $M_X$ can finally be computed through
eq.(\ref{split}) by using known ``input FFs''. They describe the
non--perturbative hadronization of quarks and gluons into mesons and
baryons, which happens at $Q = Q_0$. These FFs $D_i^h(x,Q_0^2)$, where
$i \in \{u,d,s,c,b,g\}$ and $h$ represents a hadron, can be obtained
directly from a fit to (e.g.) LEP data. We used the results of
\cite{Poetter}, where the FFs of a quark or gluon into protons,
neutrons, pions and kaons (or more exactly the sum over particles and
antiparticles) are parameterized in the form
$Nx^{\alpha}(1-x)^{\beta}$.

The original form \cite{Poetter} of these functions is only valid down
to $x = 0.1$. Kinematic and color coherence effects, which are not
included in the usual DGLAP framework, become important
\cite{Basics_of_QCD}) at $x \lsim \sqrt{(Q/Q_{\rm had})} \sim 0.1$,
where in the second step we have used the LEP energy scale $Q \sim
100$ GeV. For $Q \sim M_X \sim 10^{16}$ GeV these effects become large
only for $x \lsim 10^{-8}$; they can thus safely be ignored for many
(but not all; see below) applications.  In \cite{BarbotDrees:1} we
therefore chose a rather simple extrapolation of the functions given
in \cite{Poetter} towards small $x$. Our default choice was a $N
x^{-\alpha'}$ parameterization; $N$ and $\alpha'$ were computed by
requiring the continuity of this parameterization with the FFs of
\cite{Poetter} at some $x_0 \simeq 0.1$, energy conservation, and, as
additional constraint, an identical power law behavior at small $x$
(i.e. identical $\alpha'$) for all the FFs of a given quark into the
different hadrons. This last assumption was motivated by the fact that
we obtain such an identical power law at small $x$ during the
perturbative part of the cascade, and by the well accepted LPHD
hypothesis (Local Parton-Hadron Duality) \cite{LPHD}, which
postulates a local proportionality in phase space between the spectra
of partons and hadrons. We chose different $x_0$ for each initial
parton in such a way that we obtain $\alpha'$ between 0 and 2; the
upper bound on $\alpha'$ follows from energy conservation (the energy
integral $\int_0^1dx\,xD(x)$ has to be finite).

\begin{figure}
\setlength{\unitlength}{1cm}
\input{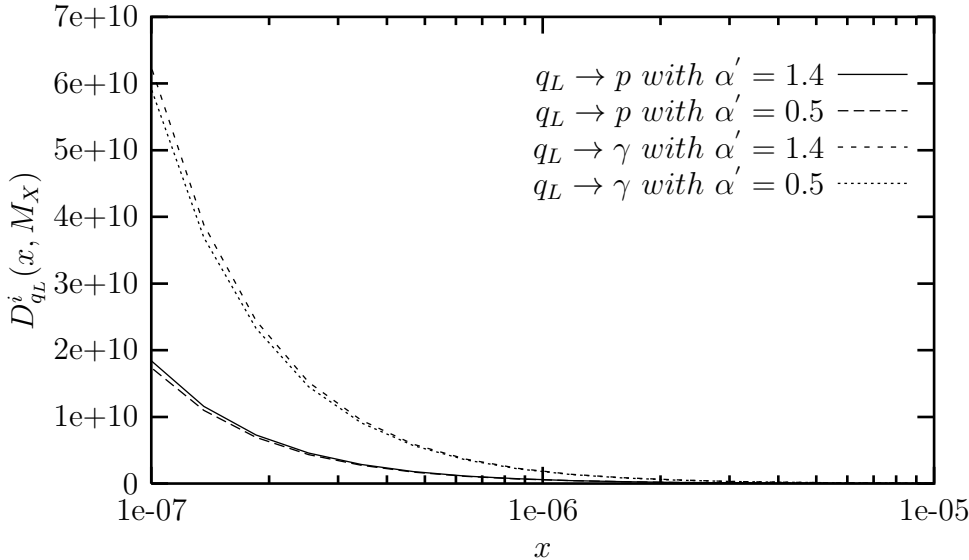}
\caption{Effect of varying the low$-x$ extrapolation of the input FFs
on the final FFs $D_{q_L}^p$ and $D_{q_L}^\gamma$. See the text for
further explanations.}
\label{smallx}
\end{figure}

In order to check the consistency of this parameterization, we used
another functional form with three free parameters: $D(x) =
ax^{-\alpha'} + b\log{x} + c\,, a > 0$. This allowed us to freely
choose $\alpha'$, keeping the same assumptions about continuity etc.
as above. This enabled us to compare two extreme values of $\alpha'$,
namely 0.5 and 1.4. The first is the smallest value compatible with $a
> 0$, while the second approximates the small$-x$ behavior of the
perturbative QCD evolution between 1 GeV and 1 TeV; requiring $\alpha'
< 1.4$ thus ensures that this perturbative evolution dominates the
behavior of the FFs at small $x$. Note also that the perhaps most
plausible value, $\alpha' \sim 1$ (which corresponds to a flat
distribution of particles in rapidity when perturbative effects are
ignored) is comfortably bracketed by these limiting values. In
fig.~\ref{smallx} we plot the final result at small $x$ for different
FFs with these two extreme parameterizations, {\it after} convolution
with the perturbative FFs. As can be seen, the effect of varying
$\alpha'$ is very small once energy conservation is imposed. This
indicates that our final results are not sensitive to the necessary
small$-x$ extrapolation of the input FFs.\footnote{However, the
original FFs of ref.\cite{Poetter} should {\em not} be used on the
whole range [$10^{-7},1$], since they violate energy conservation
badly, leading to over--production of particles at small $x$.} The
main uncertainty at moderately small $x$ ($10^{-5} \lsim x \lsim 0.1$)
will then come from perturbative higher order corrections, which might
be quite significant in this range.

Unfortunately, we were not able to perform a complete NLO analysis,
for the following reasons. Beyond leading order the SFs for
space--like and time--like processes are no longer identical
\cite{Furmanski}. Already at next--to--leading order (NLO) the
time--like SFs have a rather bad behavior at small $x$, with a
negative leading term $-\frac{40}{9}\frac{1}{x}$ in $P_{qq}$. This
term is tempered in the final spectra (which have to be positive) by
the convolution occuring in the DGLAP equations, as well as by the
convolution of the FFs with NLO ``coefficient functions'' which modify
the basic relation (\ref{def_ff}) once higher order corrections are
included. Note that the FFs, SFs and NLO coefficient functions are
scheme dependent; worse, the coefficient functions are also
process--dependent, i.e. they will depend on the spins of $X$ and its
primary decay products. NLO results are known for the classical
processes occuring in pure (non--supersymmetric) QCD, but they are not
available for most of the processes we are interested in. Moreover, in
cases where they are known, these coefficient functions often contain
the most important part of the NLO correction, rendering useless any
attempt to give a partial result by only including NLO terms in the
SFs. We conclude that it might be possible and interesting to carry
out a full NLO analysis in the pure QCD case, but this is not possible
in the more interesting supersymmetric case using available
results. Note that part of the perturbative NLO effects are absorbed
in the input FFs, through their fit to experimental data. At very
small $x$, NLO effects just give the leading ``color coherence''
corrections, which are resummed analytically in the MLLA formula, as
will be discussed in Sec.~3.4.

Finally, having computed the spectra of long--lived hadrons and
leptons, we still need to treat weak decays of unstable particles, in
order to obtain the final spectra of protons, electrons, photons and
the three types of neutrinos. This is again done using the formalism
of eq.(\ref{decay}). We limit ourselves to 2-- and 3--body decays,
considering the 4--body decays of the $\tau$ to be cascades of 2--body
decays. As before, we compute the decay functions $P_{sH}$ for $H
\rightarrow s$ decays from phase space only, and we ignore decays with
branching ratio smaller than 1\%. We then renormalize the branching
ratios of the decays we do include, so that we maintain energy
conservation. We also explicitly treated the leptonic part of the
semi--leptonic decays of $b-$ and $c-$flavored hadrons, which are
evidently not included in the FFs of \cite{Poetter}. We used the
Peterson parameterization for non--perturbative heavy quark
fragmentation \cite{Peterson}, and then treated the semi--leptonic
decays in the spectator model (i.e. using the same spectra as for free
quark decays, with $m_c = 1.5$ GeV and $m_b = 4.5$ GeV). Details of
our treatment of decays are given in Appendix C.

\section{Results and analysis}
\label{sec:results}

\subsection{General features of the final fluxes}

Some FFs computed with our code have already been presented in
ref.\cite{BarbotDrees:1}. In particular, we showed that including
supersymmetric particles in the QCD shower evolution significantly
softens the FFs of strongly interacting (s)particles. Usually FFs are
multiplied with $x^3$ to allow a more direct comparison with the
experimental UHECR spectra, which are multiplied with $E^3$ to make
them (approximately) flat. In this normalization, including
supersymmetry reduces the height of the peak of the FFs of initial
quarks and gluons by approximately a factor of two. Including
electroweak interactions reduces this peak by another 20\% or so. More
importantly, it leads to a very energetic component of the final FFs
into leptons and especially photons, if the primary particle has
electroweak interactions. This is due to the emission of very
energetic electroweak gauge bosons early in the parton shower; some of
them will decay or fragment (partly into leptonic final states), while
others survive as photons after the unitary transformation
(\ref{trafo}). In this article we focus on a detailed description of
$X$ decays computed according to the ``state of the art'', i.e. with
all interactions included.

A fairly complete set of results of our code for a given set of SUSY
parameters is given in Appendix E. Here we assumed similar masses for
all sfermions, higgsinos, heavy Higgs bosons and gluinos, $m_{\tilde
f} \simeq m_A \simeq m_{\tilde g} \simeq \mu \simeq 500$ GeV; this
leads to a gaugino--like LSP, since we assume ``gaugino mass
unification'', i.e. $6 m_{\tilde B} \simeq 3 m_{\tilde W} \simeq
m_{\tilde g}$. We also choose a small value for the ratio of vevs,
$\tan\beta = 3.6$. We see that the final spectra depend sensitively on
the primary $X$ decay products \cite{BarbotDrees:1}, especially in the
large $x$ region. This strong dependence on the unknown primary $X$
decay mode(s) should be kept in mind when one is trying to
quantitatively test ``top--down'' models. Nevertheless, we can make a
few general statements about these results. To that end we first
analyze ratios of FFs of the different stable particles divided by the
FF of the same initial particle into protons. Recall that these FFs
directly represent the flux at source if $X$ undergoes two--body
decay.

\begin{figure}
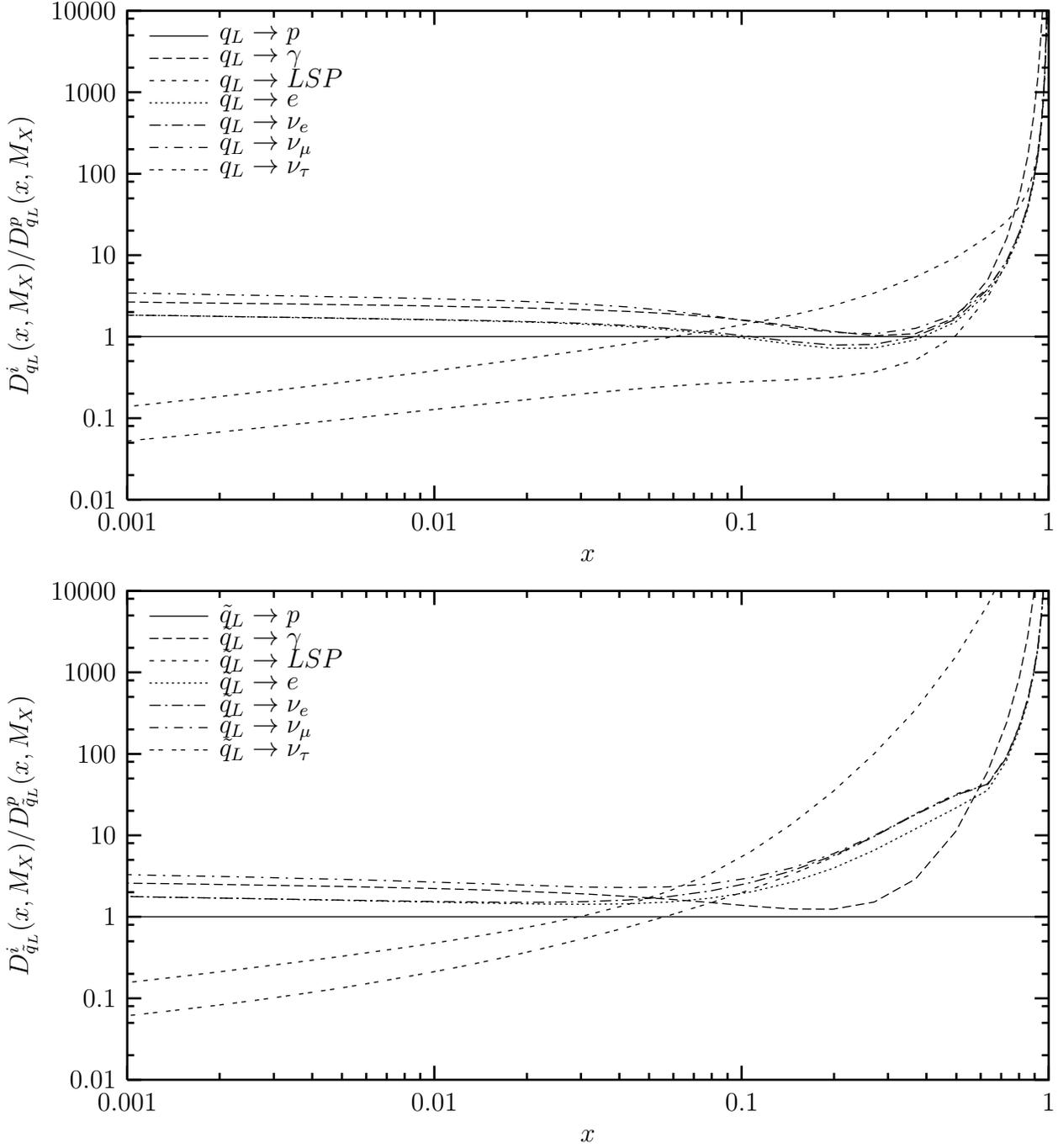

\input{Low_G_ratio_uL.tex}
\input{Low_G_ratio_uL_.tex}
\caption{Ratios of FFs $D_I^h/D_I^p$ for different stable particles $h$,
for an initial first or second generation $SU(2)$ doublet quark, $I =
q_L$, (top) or squark, $I = \tilde q_L$ (bottom).} 
\label{ratio_q}
\end{figure}

\begin{figure}[t]
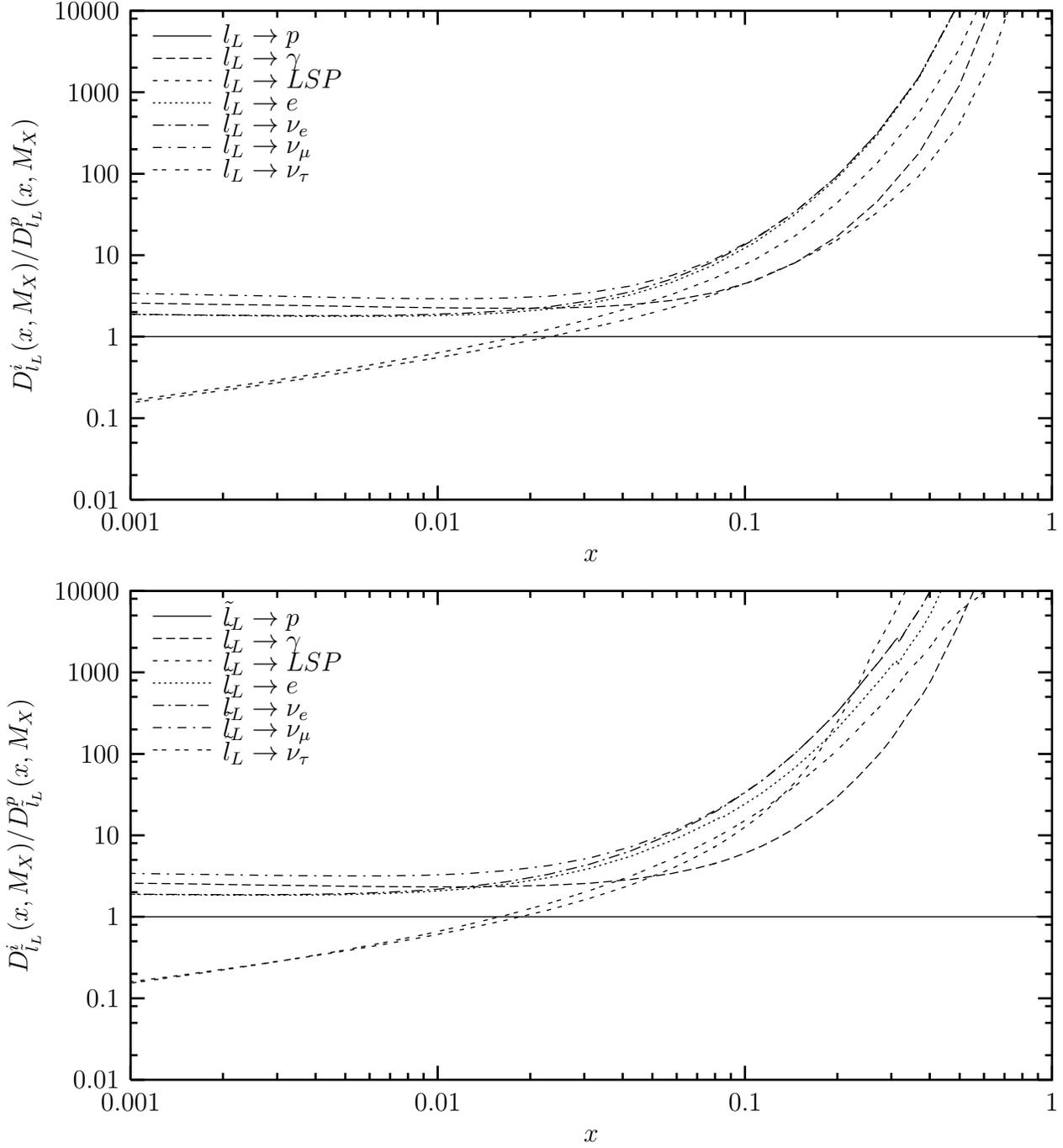

\input{Low_G_ratio_eL.tex}
\input{Low_G_ratio_eL_.tex}
\caption{As in fig.~\ref{ratio_q}, but for initial first or second
generation $SU(2)$ doublet lepton, $I = l_L$, (top) or slepton, $I =
\tilde l_L$ (bottom).}
\label{ratio_l}
\end{figure}

Taking the ratios of the different FFs renders some features more
evident, as can be seen from figs.~\ref{ratio_q} and \ref{ratio_l}.
First of all, in the low $x$ region most FFs show the same power law
behavior, and the ratios become quite independent of the initial
particle. The exceptions are the FFs into the LSP and $\nu_\tau$. This
comes from the fact that the LSP flux as well as most of the
$\nu_\tau$ are produced in the perturbative cascade above 1 TeV and in
the following decays of the heavy particles of the spectrum; they
receive no contribution from the decays of light hadrons, although the
$\nu_\tau$ flux receives a minor contribution from the decay of
$b-$flavored hadrons. In contrast, at low $x$ the fluxes of $\nu_e, \
\nu_\mu, \ e$ and $\gamma$ all dominantly originate from the decays of
light hadrons, in particular of charged or neutral pions; we saw in
fig.~\ref{smallx} that the shape of the light hadron spectrum at small
$x$ is essentially determined by the perturbative QCD evolution,
i.e. is independent of the initial particle $I$. In the region $x
\lsim 0.01$ we thus predict FFs into $\nu_\mu$ and $\gamma$ to be
approximately 3 to 4 times larger than the FF into protons, while the
FFs into electrons and $\nu_e$ are around twice the FF into
protons. The FFs into LSP and $\nu_\tau$ are five to 20 times smaller
than the one into protons. Note that the LSP flux at small $x$ from an
initial particle is almost the same as that from its superpartner. It
is determined completely by the MSSM cascade, i.e. by the
supersymmetric DGLAP equations, and is almost independent of details
of the supersymmetric spectrum. However, even at $x = 0.01$ the FF
into the LSP does retain some sensitivity to the start of the cascade,
i.e. to the initial particle $I$ and hence to the primary $X$ decay
mode(s).

At larger values of $x$ the ratios of the FFs depend more and more
strongly on the initial particle. As $x \rightarrow 1$ the proton flux
is always orders of magnitude smaller than the fluxes of all
other stable particles. One reason is that the proton is a composite
particle, i.e. its FF contains a convolution with a non--perturbative
factor which falls as a power of $1-x$ at large $x$. Even before this
convolution the flux of partons (quarks and gluons) that can give rise
to protons is suppressed at large $x$ due to copious emission of
(soft) gluons, whereas the FFs into leptons, photons and LSPs can
remain large at large $x$. If the progenitor $I$ of the cascade is a
strongly interacting superparticle, at large $x$ the FF into the LSP
always dominates over the other FFs. For an initial quark or gluon,
the flux of $\gamma$ (which is the second after LSP for a squark or
gluino) will dominate at large $x$. On the contrary, in the case of an
initial lepton, $W$, $B$ or $H_i$, the strongest fluxes will be
leptonic ones, the exact order depending of the initial
particle. Moreover, for an initial (s)lepton, the fluxes will be
significantly higher at high $x$ (and hence smaller at low $x$,
because of energy conservation) than for strongly interacting
(super)particles or Higgs bosons. Finally, an initial $B$ or $\tilde
B$ has a $\delta-$peak at $x=1$ (not visible in the figures) in
$D_B^\gamma$ and $D_{\tilde B}^{\rm LSP}$, respectively, in addition
to a smooth component that vanishes as $x \rightarrow 1$. This
behavior reflects the inability of $B$ or $\tilde B$ to radiate a
boson, i.e. there are no splitting processes $B \rightarrow B + X$ or
$\tilde B \rightarrow \tilde B + X$.

\clearpage

\clearpage

\clearpage

\noindent

\subsection{Energy distribution between the final stable particles}

In the following tables we show the total energy carried per each type
of particle at the end of the cascade, depending on the progenitor of
the cascade, for the same set of SUSY parameters as in Sec.~3.1. As
stated earlier, we are able to verify energy conservation up to at
most a few per mille at each step of the cascade, including its very
end. We see that the ``lost'' energy is somewhat larger for
(s)squarks, gluons and gluinos than for (s)leptons. This is due to
numerical artefacts. The biggest numerical uncertainties arise from
the Runge--Kutta method.\footnote{For practical reasons, we used a
fixed virtuality step in this algorithm, which we had to keep
reasonably large, the whole program being already quite
time--consuming. In the worst cases, our choice of the virtuality
step leads to errors of the order of a few per mille; such a
precision is certainly sufficient for our purposes.}

\begin{table}[t]
\begin{center}
\begin{tabular}{|c||c|c|c|c|c|c|c|c|c|c|c|c|}\hline
init (s)part $\,\,\rightarrow$
&\rule[-3mm]{0mm}{8mm} $q_L$ & $\tilde{q}_L$ & $u_R$ & $\tilde{u}_R$ &
$d_R$ & $\tilde{d}_R$ & $t_L$ & $\tilde{t}_L$ & $t_R$ & $\tilde{t}_R$
& $b_R$ & $\tilde{b}_R$ \\  
energy [\%] $\,\,\downarrow$&&&&&&&&&&&& \\ \hline
$p$ &\rule[-3mm]{0mm}{8mm} 10.0 & 8.3 & 9.1 & 7.0 & 11.5 & 8.4 & 9.3 &
8.0 & 8.8 & 7.8 & 10.3 & 8.1 \\ \hline
$\gamma$ & \rule[-3mm]{0mm}{8mm} 22.9 & 19.1 & 25.2 & 19.1 & 24.1 &
18.0 & 20.5 & 17.8 & 22.0 & 19.0 & 22.0 & 18.0 \\ \hline
$LSP$ & \rule[-3mm]{0mm}{8mm} 5.8 & 17.8 & 6.4 & 28.8 & 6.1 & 29.1 &
5.9 & 17.3 & 5.6 & 19.0 & 4.9 & 19.1 \\ \hline
$e$ & \rule[-3mm]{0mm}{8mm} 15.7 & 14.0 & 15.5 & 11.7 & 14.9 & 11.3 &
16.5 & 14.5 & 16.4 & 13.9 & 16.3 & 14.1 \\ \hline
$\nu_e$ & \rule[-3mm]{0mm}{8mm} 15.6 & 14.0 & 15.2 & 11.5 & 14.7 &
11.2 & 16.4 & 14.5 & 16.2 & 13.8 & 16.1 & 13.9 \\ \hline
$\nu_\mu$ & \rule[-3mm]{0mm}{8mm} 28.0 & 24.2 & 27.5 & 20.8 & 27.8 &
20.9 & 27.5 & 24.1 & 26.9 & 23.2 & 27.9 & 23.5 \\ \hline 
$\nu_\tau$ & \rule[-3mm]{0mm}{8mm} 1.3 & 1.8 & 0.4 & 0.4 & 0.3 & 0.4 &
3.0 & 2.9 & 3.4 & 2.6 & 1.6 & 2.3 \\ \hline \hline
sum & \rule[-3mm]{0mm}{8mm} 99.2 & 99.2 & 99.3 & 99.3 & 99.2 & 99.2 &
99.2 & 99.2 & 99.2 & 99.2 & 99.1 & 99.2 \\ \hline
\end{tabular}
\caption{Energy fractions $\int_0^1 dx \, x D_I^p(x,M_X^2)$ carried by
the stable particles $p$ at the end of the cascade, for initial
(s)quarks of the 1st/2nd and 3rd generations. We took $M_X = 10^{16}$
GeV and a sparticle spectrum with gaugino--like LSP, as described in
Sec.~3.1.} 
\label{nrj_quarks}
\end{center}
\end{table}

Note that even for an initial quark or gluon, more than 35\% of the
energy is carried by the electromagnetic channels (electrons plus
photons), while neutrinos carry about 40\%; in this case most of these
fluxes originate from the decays of light hadrons, chiefly pions. The
corresponding numbers for superparticles are slightly smaller, the
difference being made up by the increased energy fraction carried by the
LSP (at large $x$); an initial $SU(2)$ singlet squark leads to a
higher energy fraction in LSPs, since $SU(2)$ singlet sfermions
usually decay directly into the LSP, which is Bino--like for our
choice of parameters, whereas $SU(2)$ doublet sfermions preferentially
decay via a cascade involving $\tilde \chi_2^0$ or $\tilde
\chi_1^\pm$. 

Lepton--induced showers have a far smaller photon component, but now
an even larger fraction of the energy is carried by electrons and/or
neutrinos, while protons carry at most 2\%
\clearpage

\begin{table}[t]
\begin{center}
\begin{tabular}{|c||c|c|c|c|c|c|c|c|}\hline
initial (s)particle $\,\,\rightarrow$
&\rule[-3mm]{0mm}{8mm} $l_L$ & $\tilde{l}_L$ & $e_R$ & $\tilde{e}_R$ &
$\tau_L$ & $\tilde{\tau}_L$ & $\tau_R$ & $\tilde{\tau}_R$ \\
energy fraction (in \%)$\,\, \downarrow$&&&&&&&& \\ \hline
$p$ & \rule[-3mm]{0mm}{8mm} 1.2 & 2.2 & 0.1 & 0.1 & 1.2 & 2.1 & 0.1 &
1.0 \\ \hline
$\gamma$ & \rule[-3mm]{0mm}{8mm} 4.5 & 6.4 & 6.1 & 5.1 & 10.4 & 9.6 &
20.0 & 11.3 \\ \hline
$LSP$ & \rule[-3mm]{0mm}{8mm} 2.6 & 28.5 & 2.0 & 47.6 & 2.7 & 30.5 &
1.8 & 36.6 \\ \hline
$e$ & \rule[-3mm]{0mm}{8mm} 29.6 & 19.2 & 60.2 & 31.0 & 9.1 & 8.9 &
14.3 & 7.1 \\ \hline
$\nu_e$ & \rule[-3mm]{0mm}{8mm} 29.6 & 19.1 & 15.2 & 7.9 & 9.1 & 8.8 &
14.1 & 6.9 \\ \hline
$\nu_\mu$ & \rule[-3mm]{0mm}{8mm} 31.1 & 21.9 & 15.3 & 8.0 & 12.9 &
12.8 & 19.5 & 9.8 \\ \hline
$\nu_\tau$ & \rule[-3mm]{0mm}{8mm} 1.1 & 2.2 & 0.1 & 0.1 & 54.3 & 27.1
& 30.0 & 27.1 \\ \hline \hline
sum\rule[-3mm]{0mm}{8mm} & 99.8 & 99.7 & 99.8 & 99.8 & 99.8 & 99.7 &
99.8 & 99.7 \\ \hline
\end{tabular}
\caption{Energy fractions carried by the stable particles at the end
of the cascade, for initial (s)leptons of the 1st/2nd and 3rd
generations. Parameters are as in Table~1.}
\label{nrj_leptons}
\end{center}
\end{table}
\begin{center}
\begin{table}
\begin{center}
\begin{tabular}{|c||c|c|c|c|c|c|c|c|c|c|}\hline
initial (s)particle $\,\,\rightarrow$
& \rule[-3mm]{0mm}{8mm} $B$ & $\tilde{B}$ & $W$ & $\tilde{W}$ & $g$ &
$\tilde{g}$ & $H_1$ & $\tilde{H}_1$ & $H_2$ & $\tilde{H}_2$ \\
energy fraction (in \%) $\,\,\downarrow$ &&&&&&&&&& \\ \hline
$p$ & \rule[-3mm]{0mm}{8mm} 1.8 & 1.5 & 7.3 & 6.1 & 9.8 & 9.1 & 8.5 &
7.0 & 8.0 & 5.6 \\ \hline
$\gamma$ & \rule[-3mm]{0mm}{8mm} 71.6 & 4.1 & 17.8 & 14.2 & 22.5 &
20.7 & 19.4 & 16.7 & 18.9 & 14.1 \\ \hline
$LSP$ & \rule[-3mm]{0mm}{8mm} 4.2 & 76.9 & 7.0 & 24.5 & 8.4 & 14.0 &
4.9 & 18.6 & 4.9 & 31.2 \\ \hline
$e$ & \rule[-3mm]{0mm}{8mm} 7.2 & 5.7 & 17.0 & 14.0 & 15.2 & 14.4 &
17.2 & 14.6 & 17.1 & 12.4 \\ \hline
$\nu_e$ & \rule[-3mm]{0mm}{8mm} 5.2 & 4.0 & 17.4 & 14.1 & 15.0 & 14.2
& 17.2 & 14.6 & 17.5 & 12.5 \\ \hline
$\nu_\mu$ & \rule[-3mm]{0mm}{8mm} 7.6 & 5.9 & 26.4 & 21.6 & 27.1 &
25.4 & 27.2 & 22.9 & 27.1 & 19.3 \\ \hline
$\nu_\tau$ & \rule[-3mm]{0mm}{8mm} 2.1 & 1.7 & 6.5 & 4.9 & 1.0 & 1.2 &
5.1 & 4.4 & 6.1 & 4.2 \\ \hline \hline
sum & \rule[-3mm]{0mm}{8mm} 99.8 & 99.8 & 99.4 & 99.4 & 99.1 & 98.9 &
99.5 & 98.9 & 99.5 & 99.3 \\ \hline
\end{tabular}
\caption{Energy fractions carried by the stable particles at the end
of the cascade, for initial bosons and bosinos. Parameters are as in
Table~1.}
\label{nrj_bosons}
\end{center}
\end{table}
\end{center}
\clearpage

\noindent
of the primary's energy. In this case the difference between an
initial particle and its superpartner is much larger than in case of
strongly interacting particles, since a much higher fraction of an
initial slepton's energy goes into LSPs, due to the reduced
perturbative shower and shorter superparticle decay cascades. This
also explains why more than 70\% of the energy of an initial $B$
($\tilde B$) goes into photons (LSPs).  The energy fractions for an
initial $SU(2)$ gauge or Higgs boson resemble those for a quark (with
the exception of an increased $\nu_\tau$ component, which is however
washed out by neutrino oscillations), although the shapes of the
corresponding FFs differ quite dramatically. The energy fraction
carried by protons is always quite small. Pions are created much more
abundantly in the non--perturbative hadronization, and decay into
leptons (2/3) and photons (1/3). As noted earlier, this explains the
regularity and the features of the small $x$ behavior.

\subsection{Dependence on SUSY parameters}
\label{subsec:SUSY_dep}

As stated in \cite{BarbotDrees:1}, the general features of our results
described above depend very little on the set of SUSY parameters we
are using. Here we give a more precise analysis of the influence of
different parameters describing the SUSY spectrum. As usual we present
our results as $x^3 \cdot D_I^p(x,M_X^2)$. The multiplication with
the third power of the energy leads to an approximately flat cosmic
ray spectrum for $E \lsim 10^{10}$ GeV \cite{reviewSigl}. In our case
it suppresses the small$-x$ region, leading to maxima in the curves at
$x$ between 0.1 and 1.

We first studied the dependence of our results on the overall SUSY
mass scale, by comparing results for two different ISASUSY input mass
scales for scalars and gluinos: $M_{SUSY} \sim 500$ GeV and $1$
TeV. As expected, this change has almost no impact on the final
results, since the details of the decay chains of heavy (s)particles
will depend mostly on the relative ordering of the (s)particle spectrum
(e.g. allowing or preventing some decay modes), rather than on their
absolute mass scale. Moreover, a factor 2 or 3 in the scale where the
MSSM evolution is terminated does not change the FFs much, since the
DGLAP equations describe an evolution which is only logarithmic in the
virtuality.

\begin{figure}[ht]
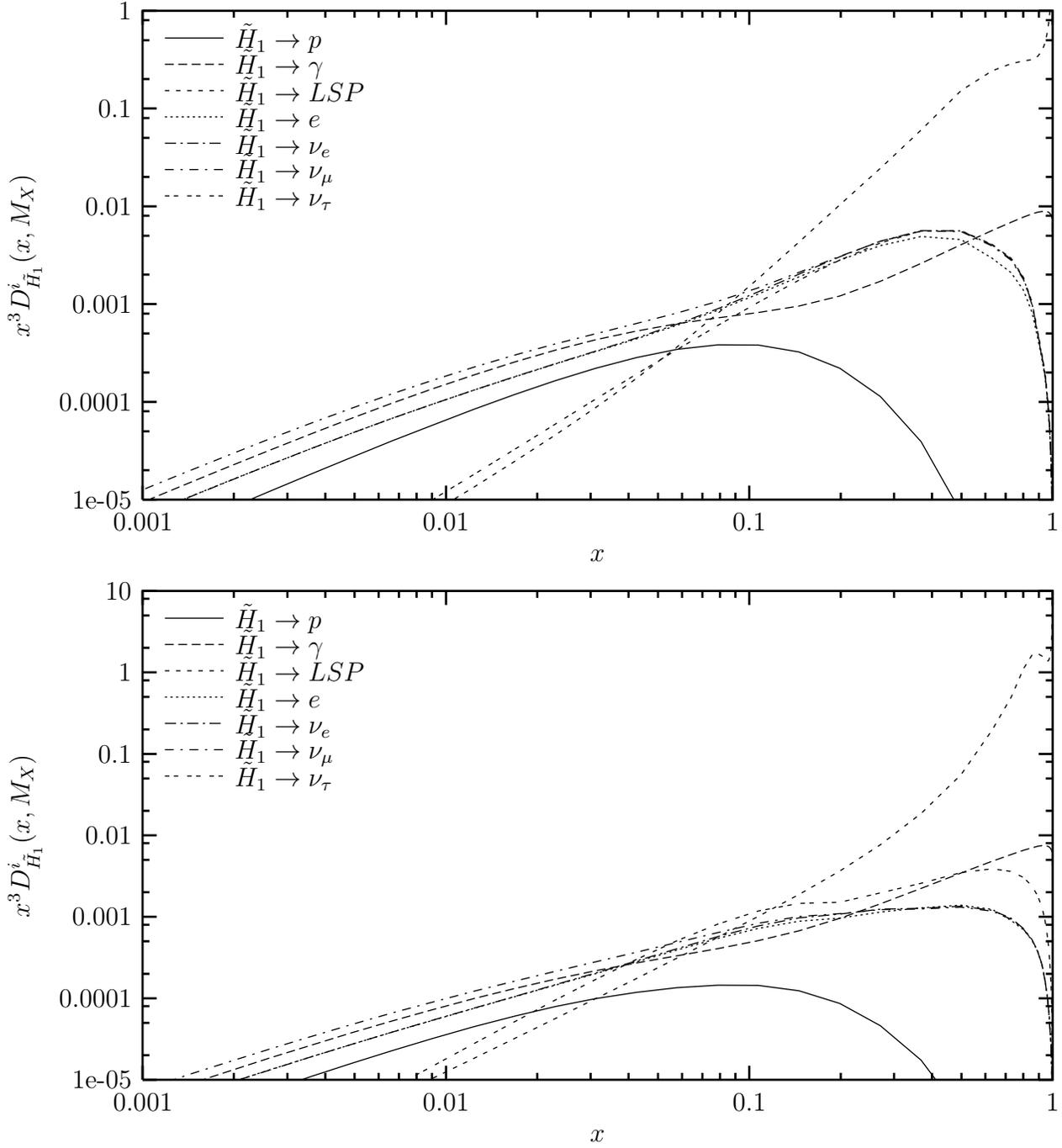

\input{Low_H_H1_.tex}
\input{Big_H_H1_.tex}
\caption{FFs into the final stable particles for an initial
$\tilde{H}_1$ for $\tan\beta = 3.6$ (top) or 48 (bottom).}
\label{LB1}
\end{figure}

Next we compared two rather extreme values of $\tan \beta$, namely 3.6
and 48, leaving all dimensionful parameters at the weak scale
unchanged. Once again the effect is very small. The only visible
difference occurs for initial $H_1$ and $\tilde{H}_1$, where the
increase of $\tan \beta$ produces more $\nu_\tau$ at large $x$, as can
be seen in fig.~\ref{LB1}. However, flavor oscillations will
essentially average the three neutrino fluxes between source and
detector, so we expect very little direct dependence of measurable
quantities on $\tan\beta$. The main remaining effect is an increase of
the overall multiplicity by $\sim 30\%$ for an initial $H_1$ or
$\tilde H_1$ in case of large $\tan\beta$, due to the increased shower
activity from the much larger bottom Yukawa coupling. However, the
situation could be different in more constrained models, where the
spectrum is described by a few soft breaking parameters specified at
some high energy scale. In this case a change of $\tan\beta$ generally
changes the sparticle and Higgs spectrum, and can also greatly modify
some branching ratios.

In order to get a feeling for how the various FFs depend on the
relative ordering of the dimensionful parameters describing the SUSY
spectrum, we investigated two rather extreme cases. They resemble two
qualitatively different regions of parameter space in the minimal su-

\clearpage

\clearpage

\noindent
pergravity (mSUGRA or CMSSM) model where the thermal LSP relic density
is acceptably small \cite{msugra}.\footnote{In our case $X$ particles
could contribute significantly to the Dark Matter; in this scenario,
which is realized only for a small region of the total allowed $M_X,
\tau_X$ plane, the upper bound on the LSP relic density would have to
be tightened accordingly, but the allowed regions of parameter space
would be qualitatively the same.} In the first scenario the LSP
$\tilde \chi_1^0$ has small mass splitting to the lightest stau,
\stauone. We took the following values for the relevant soft breaking
parameters: $m_{\tilde q} \simeq m_{\tilde g} = 1$ TeV for all
squarks, $m_{\tilde l_L} = 250$ GeV for all $SU(2)$ doublet sleptons,
$m_{\tilde l_R} \simeq 200$ GeV for $l = e, \, \mu$ but reduced
$m_{\tilde \tau_R}$ so that $m_{\tilde \tau_1} = m_{\tilde \chi_1^0} +
13$ GeV $= 163$ GeV; note that in mSUGRA one needs large mass
splitting between squarks and sleptons if the LSP mass is to be close
to the \stauone\ mass. The physical sfermion masses receive additional
contributions from $SU(2) \times U(1)_Y$ symmetry breaking, and, in
case of the third generation, from mixing between singlet and doublet
sfermions; in case of $\tilde t$, contributions $+m_t^2$ to the
diagonal entries of the mass matrix also have to be added.  Our choice
$\mu = 1$ TeV together with the assumption of gaugino mass unification
ensures that the LSP is an almost pure bino.

In contrast, in the second scenario we took $\mu = -100$ GeV,
$m_{\tilde g} = 800$ GeV, so that the LSP is dominated by its higgsino
components, although the bino component still contributes $\sim
20\%$. In this scenario we took $m_{\tilde q} = 1.5$ TeV for all
squarks and $m_{\tilde l} = 1.2$ TeV for all sleptons, since in mSUGRA
large scalar masses are required if the LSP is to have a large
higgsino component. We took CP--odd Higgs boson mass $m_A = 1$ TeV in
both cases, and $\tan\beta = 3.6$; we just saw that the latter choice
is not important for us. In the following we will refer to these two
choices as the ``gaugino'' and ``higgsino'' set of parameters,
respectively.

\begin{figure}
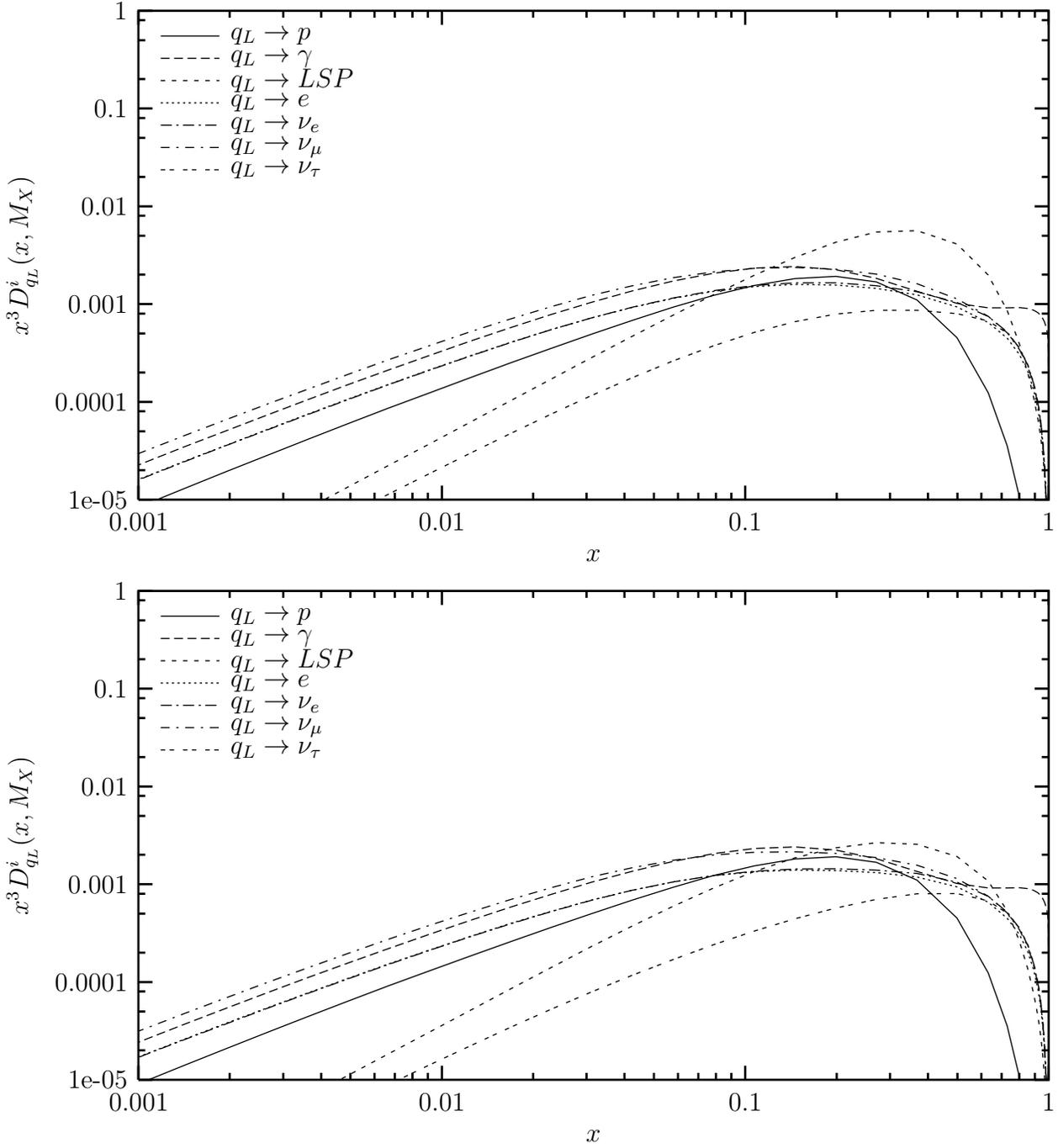

\input{Low_Ga_uL.tex}
\input{Low_Hi_uL.tex}
\caption{FFs into the final stable particles for an initial
$SU(2)$ doublet quark of the first or second generation $q_L$, for
the gaugino (top) and higgsino (bottom) set of parameters.}
\label{GH_uL}
\end{figure}

In Fig.~\ref{GH_uL} we compare the FFs of an initial first or second
generation $SU(2)$ doublet quark $q_L$ for these two scenarios. The
main difference occurs in the FF into the LSP, which is significantly
softer for the higgsino set. The reason is that most heavy
superparticles (sfermions and gluinos) preferentially decay into
gaugino--like charginos and neutralinos, which have much larger
couplings to most squarks than the higgsino--like states do. These
gaugino--like states are the lighter two neutralinos and lighter
chargino in case of the gaugino set, but they are the heavier $\tilde
\chi$ states for the higgsino set. The supersymmetric decay chains
therefore tend to be longer for the higgsino set, which means that
less energy goes into the LSP produced at the very end of each chain.

\begin{figure}
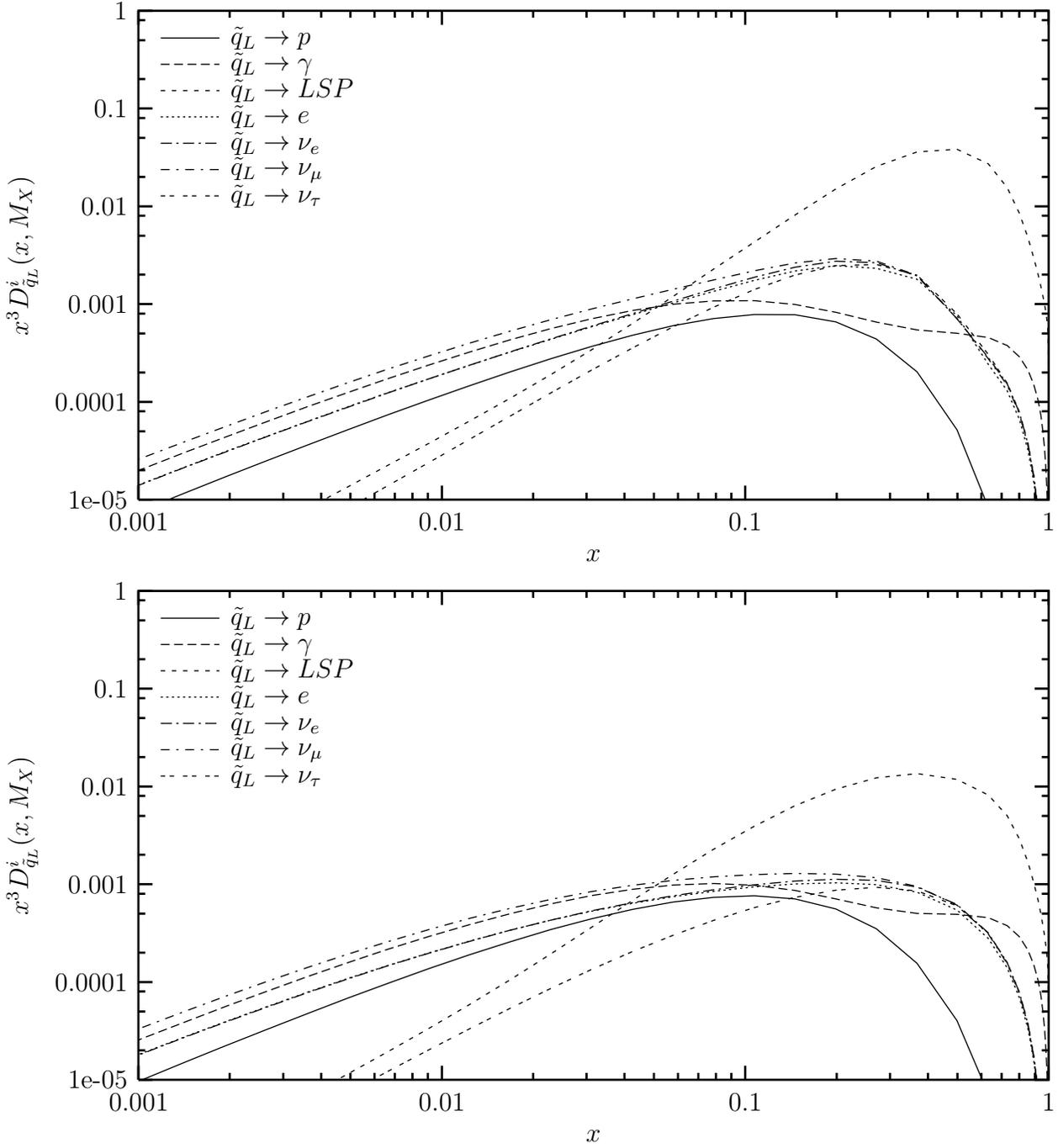

\input{Low_Ga_uL_.tex}
\input{Low_Hi_uL_.tex}
\caption{FFs into the final stable particles for an initial
$SU(2)$ doublet squark of the first or second generation $\tilde q_L$,
for the gaugino (top) and higgsino (bottom) set of parameters.}
\label{GH_uL_}
\end{figure}

Fig.~\ref{GH_uL_} shows the same comparison for an initial first or
second generation $SU(2)$ doublet squark $\tilde q_L$. Not
surprisingly, the FFs of a squark are more sensitive to details of the
sparticle spectrum than those of a quark. In particular, in addition
to the reduced FF into the LSP, we now also see that the FFs into
neutrinos and electrons are suppressed for the higgsino set relative to
the gaugino set. This is partly again due to the longer decay chains,
which pushes these FFs towards smaller $x$ where the $x^3$
normalization factor suppresses them more strongly, and partly because
the branching ratios for leptonic decays of the $SU(2)$ gaugino--like
$\tilde \chi$ states are smaller here than for the gaugino set, which
implies that fewer leptons are produced in sparticle decays. On the
other hand, the longer decay chains and larger hadronic branching
ratios for $\tilde \chi$ decays characteristic of the higgsino set
lead to an increase of the total multiplicity of 25\% or so, as can be
seen from the FFs at small $x$; of course, in this region the ratios
of these FFs again approach their universal values, as discussed in
Sec.~3.1.

\clearpage

\clearpage

\clearpage

\begin{figure}
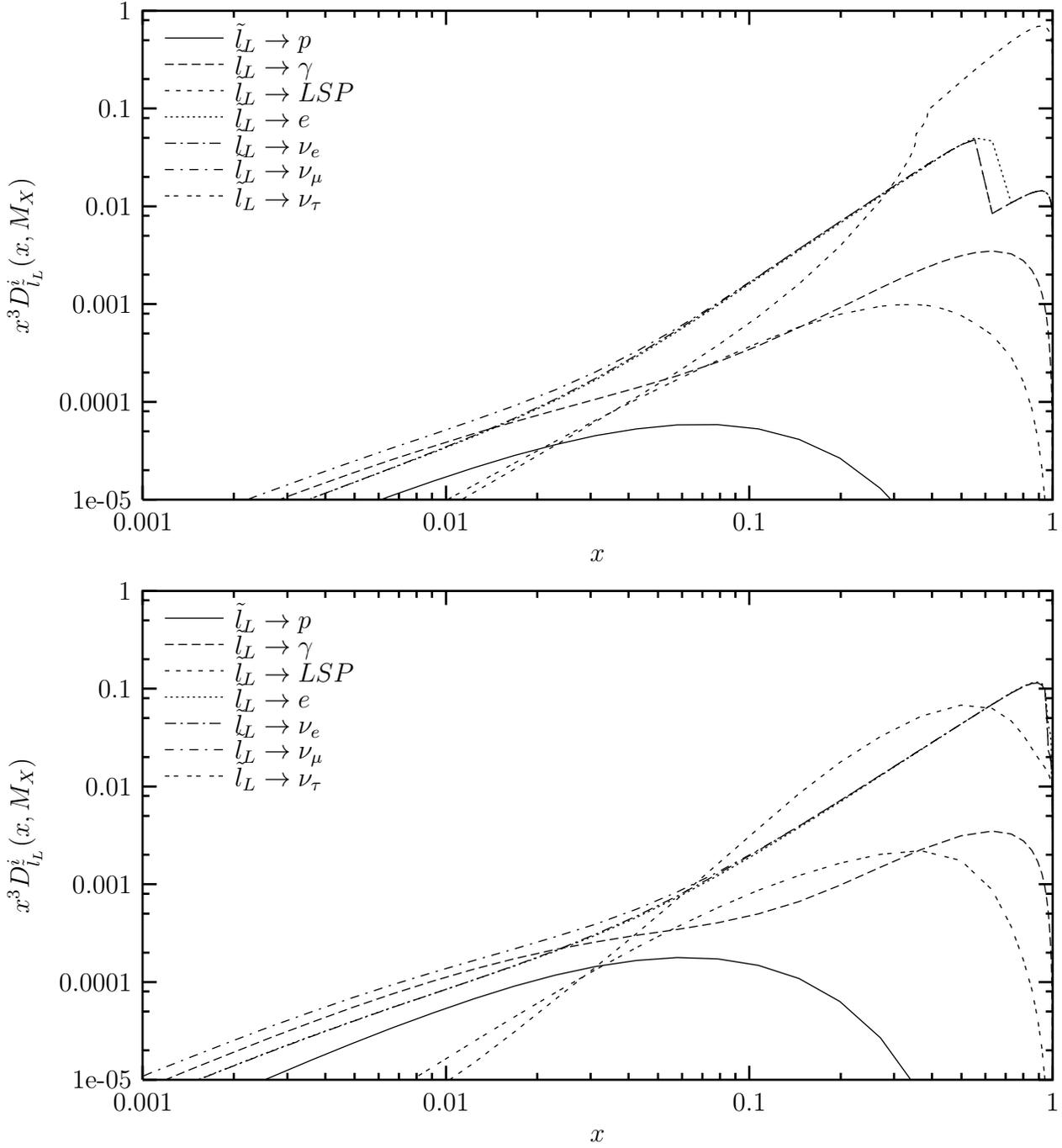

\input{Low_Ga_eL_.tex}
\input{Low_Hi_eL_.tex}
\caption{FFs into the final stable particles for an initial
$SU(2)$ doublet slepton of the first or second generation $\tilde l_L$,
for the gaugino (top) and higgsino (bottom) set of parameters.}
\label{GH_eL_}
\end{figure}

If the initial particle is strongly interacting, the rapid evolution
of the shower ensures that the generalized FFs (\ref{boundary_MSSM})
describing the evolution between $M_{\rm SUSY}$ and $M_X$ essentially
vanish at $x \simeq 1$, i.e. all spectra are smooth. In contrast, if
the initial particle $I$ has only weak interactions, a significant
$\delta-$peak will remain at $x = 1$ in the generalized FF $\tilde
D_I^I$. If $I$ is a superparticle or Higgs boson, the decays of $I$
can therefore lead to sharp edges in the final FFs. This is
illustrated in Fig.~\ref{GH_eL_}, which shows the FFs for an initial
first or second generation $SU(2)$ doublet sleptons $\tilde l_L$. The
parameters of the gaugino set are chosen such that $\tilde l_L$
sleptons can only decay into $l \ +$ LSP. The decays of the $\tilde
l_L$ which survive at $x=1$ therefore lead to edges in the FFs into
$e, \, \nu_e$ and $\nu_\mu$; recall that $\tilde l_L$ is an equal
mixture of $\tilde e_L, \, \tilde \nu_e, \, \tilde \mu_L$ and $\tilde
\nu_\mu$. The edge in the FF into $e$ occurs at a somewhat larger
value of $x$ than those in the FFs into $\nu_{e,\mu}$, since after
$SU(2)$ symmetry breaking the charged members of the slepton doublets
are a little heavier than the neutral ones; the decay $\tilde e_L
\rightarrow e \tilde \chi_1^0$ therefore deposits more energy in the
electron than $\tilde \nu_e \rightarrow \nu_e \tilde \chi_1^0$
deposits in the neutrino. However, in both cases the bulk of the
energy goes into the LSP, which is rather close in mass to the
slepton. This is quite different for the higgsino set, where sleptons
are much heavier than all $\tilde \chi$ states. As a result, almost
the entire slepton energy can go into the decay lepton, leading to FFs
into $e, \, \nu_e$ and $\nu_\mu$ that are peaked very near $x=1$
(after multiplying with $x^3$). Furthermore, since most sleptons now
first decay into heavier $\tilde \chi$ states rather than directly
into $\tilde \chi_1^0$, the FF into the LSP is much softer than for
the gaugino set. Finally, the effect of the longer decay chains of
SUSY particles on the overall multiplicity now amounts to about a
factor of 2, and is thus much more pronounced than for initial
squarks; this can be explained by the reduced importance of the shower
evolution in case of only weakly interacting primaries.

Fig.~\ref{GH_H1} shows that in case of an initial $H_1$ Higgs doublet,
the role of the two parameter sets is in some sense reversed. Recall
that we chose $\tan\beta > 1$ and $m_A \gg M_Z$. In that case the
heavy Higgs bosons mostly consist of various components of the $H_1$
doublet, with only small admixtures of $H_2$; see eq.(\ref{smtrafo})
in Appendix B. As usual with only weakly interacting primaries,
the generalized FF $D_{H_1}^{H_1}$ remains sizable at $x=1$ even at
scale $M_X$. In the higgsino set, the dominant decay modes of the
heavy Higgs bosons involve a gaugino and a higgsino, leading to a
large FF into the LSP in this case. Since in the gaugino set the mass
of the higgsino--like $\tilde \chi$ states is very close to the mass
of the heavy Higgs bosons, these supersymmetric decay modes are closed
for the heavy Higgs bosons in this case, which instead predominantly
decay into top quarks, with decays into $b$ quarks and $\tau$ leptons
also playing some role. The fragmentation and decay products of these
heavy quarks lead to a significantly larger FF into protons in the
gaugino region; semi--leptonic $t$ and $b$ decays as well as the
$\tau$ decays also lead to enhanced FFs into electrons and neutrinos
for the gaugino set. Finally, the hadronic showers initiated by the
decay products of the top quarks as well as by the $b$ quarks produced
directly in the decays of Higgs bosons raise the total multiplicity
for the gaugino set to a value which is slightly larger than that for
the higgsino set.

As final example we compare the FFs of an initial $\tilde H_2$
higgsino doublet in Fig.~\ref{GH_H2_}. Here we again find a larger FF
into the LSP for the higgsino set, including a peak at $x=1$. In this
case this is simply a reflection of the large $\tilde H_2^0$ component
of the LSP. On the other hand, in case of the gaugino set $\tilde H_2$
projects almost exclusively into the heavier $\tilde \chi$ states,
which have many two-

\clearpage

\clearpage

\begin{figure}
\input{Low_Ga_H1.tex}
\input{Low_Hi_H1.tex}
\caption{FFs into the final stable particles for an initial $H_1$
Higgs doublet, for the gaugino (top) and higgsino (bottom) set of
parameters.}
\label{GH_H1}
\end{figure}

\clearpage
\begin{figure}
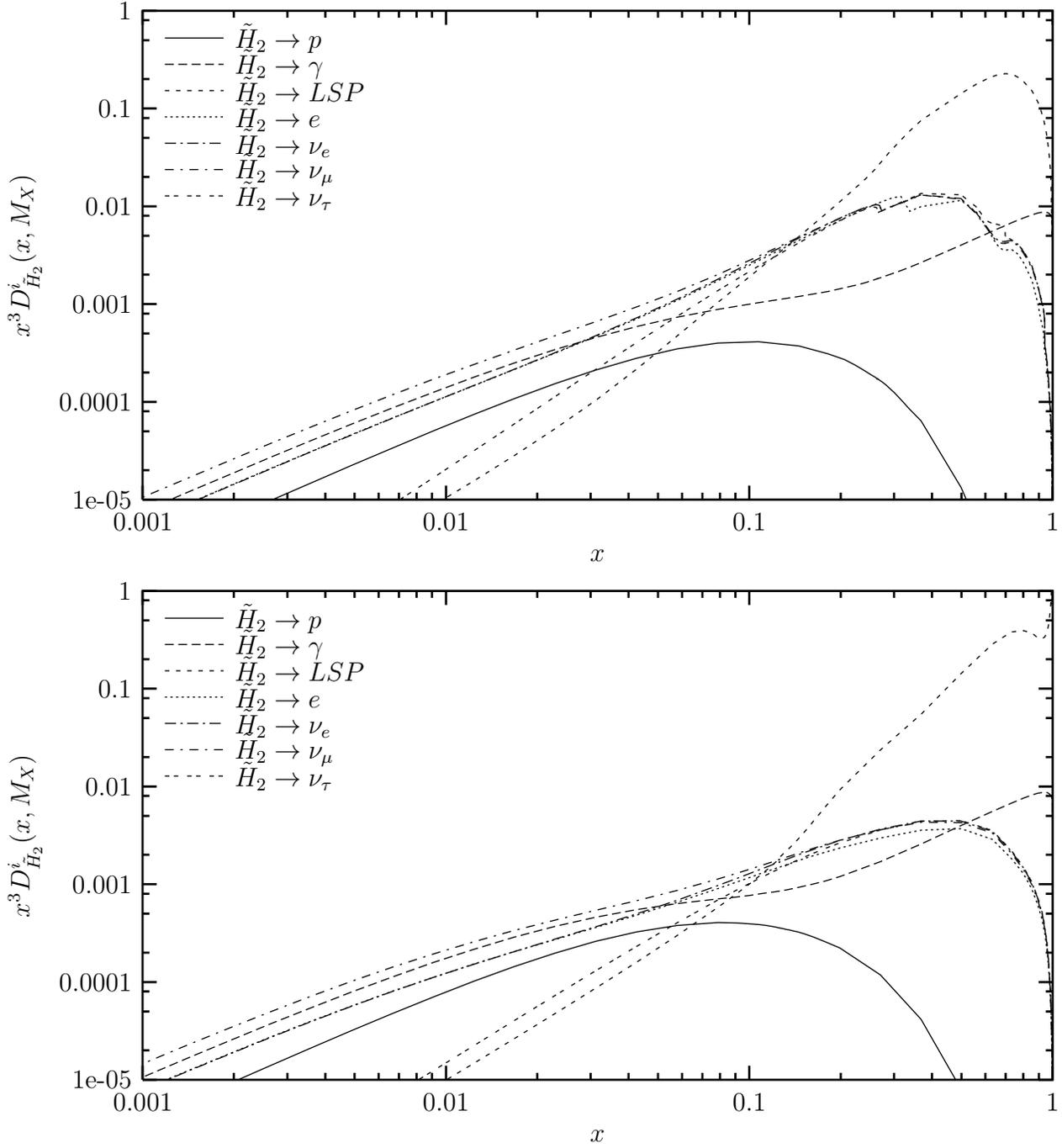

\input{Low_Ga_H2_.tex}
\input{Low_Hi_H2_.tex}
\caption{FFs into the final stable particles for an initial $\tilde H_2$
higgsino doublet, for the gaugino (top) and higgsino (bottom) set of
parameters.}
\label{GH_H2_}
\end{figure}

\clearpage

\noindent
body decay modes into sleptons and leptons. This explains the relative
enhancement at large $x$ of the FFs into leptons that we observe for
the gaugino set, as well as the structures in these FFs. On the other
hand, the longer sparticle decay chains again imply a somewhat larger
overall multiplicity for the higgsino set. These decays of heavy
sparticles are important here since the large top Yukawa coupling of
$\tilde H_2$ initiates a significant parton shower in this case, where
numerous superparticles are produced. This is quite different for an
initial $\tilde H_1$ at small $\tan\beta$ (not shown), where we find a
{\em smaller} overall multiplicity for the higgsino set, since the
number of produced superparticles remains small, and the initial
particle $\tilde H_1$ has a longer decay chain for the gaugino set.

Altogether we see that the SUSY spectrum can change the final FFs, and
thus the final spectra of $X$ decay products, significantly. Generally
this effect is stronger for an initial superparticle or heavy Higgs
boson than for an SM particle, and stronger for only weakly
interacting particles than for those with strong interactions.
However, with the exception of the FFs into the LSP, the variation is
usually not more than a factor of two, and often much less. The
dependence of the $X$ decay spectra on SUSY parameters can therefore
be significant for detailed quantitative analyses, but this dependence
is always weaker than the dependence on the primary $X$ decay mode(s).

\subsection{Coherence effects at small $x$: the MLLA solution}
\label{sec:MLLA}

So far we have used a simple power law extrapolation of the hadronic
(non--perturbative) FFs at small $x$. This was necessary since the
original input FFs of ref.\cite{Poetter} are valid only for $x \geq
0.1$. As noted earlier, we expect our treatment to give a reasonable
description at least for a range of $x$ below 0.1. However, at very
small $x$, color coherence effects should become important
\cite{Basics_of_QCD}. These lead to a flattening of the FFs, giving a
plateau in $x D(x)$ at $x_{\rm plateau} \sim \sqrt{Q_{\rm had}/M_X}
\sim 10^{-8}$ for $M_X = 10^{16}$ GeV. One occasionally needs the FFs
at such very small $x$. For example, the neutrino flux from $X$ decays
begins to dominate the atmospheric neutrino background at $E \sim
10^5$ GeV \cite{neutrinos, bdhh1}, corresponding to $x \sim 10^{-11}$
for our standard choice $M_X \sim 10^{16}$ GeV. In this subsection we
therefore describe a simple method to model color coherence effects in
our FFs.

This is done with the help of the so--called limiting spectrum derived
in the modified leading log approximation. The key difference to the
usual leading log approximation described by the DGLAP equations is
that QCD branching processes are ordered not towards smaller
virtualities of the particles in the shower, but towards smaller
emission angles of the emitted gluons; note that gluon radiation off
gluons is the by far most common radiation process in a QCD
shower. This angular ordering is due to color coherence, which in the
conventional scheme begins to make itself felt only in NLO (where the
emission of two gluons in one step is treated explicitly). It changes
the kinematics of the parton shower significantly. In particular, the
requirement that emitted gluons still have sufficient energy to form
hadrons strongly affects the FFs at small $x$. For sufficiently high
initial shower scale and sufficiently small $x$ the MLLA evolution
equations can be solved explicitly in terms of a one--dimensional
integral \cite{Basics_of_QCD}. This essentially yields the modified FF
describing the perturbative gluon to gluon fragmentation, $\tilde D_g^g$
in the language of eq.(\ref{split}). In order to make contact with
experiment, one makes the additional assumption that the FFs into
hadrons coincide with $\tilde D_g^g$, up to an unknown constant; this
goes under the name of ``local parton--hadron duality'' (LPHD)
\cite{LPHD}. Here we use the fit of this ``limiting spectrum'' in
terms of a distorted Gaussian \cite{distorted_gaussian}, which
(curiously enough) seems to describe LEP data on hadronic FFs somewhat
better than the ``exact'' MLLA prediction does. It is given by
\beq \label{gaussian}
F_i(\xi,\tau) \equiv xD_i(x,Q) =
\frac {\bar{n}_i} {\sigma \sqrt{2\pi} } \exp {\left[ \frac {1} {8} k +
\frac{1}{2} s \delta - \frac{1}{4} (2+k) \delta^2 + \frac{1}{6} s
\delta^3 + \frac{1}{24} k \delta^4 \right]},
\eeq
where $\bar{n}_i$ is the average multiplicity. The other quantities
appearing in eq.(\ref{gaussian}) are defined as follows:
\beqa \label{gauss_coef}
\tau &=& \log{\frac{Q}{\Lambda}} \,,\nonumber\\
\xi &=& \log{\frac{1}{x}} \,,\nonumber\\
\bar{\xi} &=& \frac{1}{2} \tau \left( 1 + \frac {\rho} {24}
\sqrt{\frac{48}{\beta \tau} } \right) + {\cal O}(1)\,,\nonumber\\
\sigma &=& \langle (\xi - \bar{\xi})^2 \rangle^{1/2} =
\sqrt{ \frac{1}{3} } \left( \frac{\beta}{48} \right)^{1/4} \tau^{3/4}
\left( 1 - \frac{1}{64} \sqrt{ \frac{48\beta} {\tau} } \right) + {\cal
O} (\tau^{-1/4}) \,,\nonumber\\
\delta &=& \frac{ \xi -\bar{\xi}} {\sigma} \,,\nonumber\\
s &=& \frac{ \langle (\xi - \bar{\xi})^3 \rangle }{\sigma^3} =
-\frac {\rho}{16} \sqrt{ \frac{3} {\tau} } \left( \frac {48}
{\beta\tau} \right)^{1/4} + {\cal O}(\tau^{-5/4}) \,,\nonumber\\
k &=& \frac{ \langle (\xi - \bar{\xi})^4 \rangle }{\sigma^4} =
- \frac {27} {5\tau} \left( \sqrt{ \frac{1}{48} \beta \tau } - \frac
{1}{24} \beta \right) + {\cal O}(\tau^{-3/2}) \,.
\eeqa
where $\beta$ is the coefficient in the one--loop beta--function of
QCD and $\rho = 11 + 2 N_f/27$, $N_f$ being the number of active
flavors. Eqs.(\ref{gaussian}) and (\ref{gauss_coef}) have been derived
in the SM, where $\beta = 11 - 2 N_f/3$. Following
ref.\cite{limiting_spectrum} we assume that it remains valid in the
MSSM, with $\beta = 3$ above the SUSY threshold $M_{\rm SUSY}$ and
$\rho = 11 + 8/9$. Note that we do not attempt to model the transition
from the full MSSM to standard QCD here; indeed, we do not know of an
easy way to do this, since the limiting spectrum cannot be written as
a convolution of two other spectra. On the other hand, the position
$\bar\xi$ of the plateau depends only on $\sqrt{\beta}$, and only via
the second term, which is suppressed by a factor $\sqrt{\tau} \sim
6.5$, whereas the parameters $\sigma$ and $s$ describing the behavior
in the vicinity of the maximum depend in leading order in $\tau$ only
on $\beta^{1/4}$. Finally, the coefficient $\rho$ is very similar in
the SM and MSSM. We therefore expect the error we make by ignoring the
transition from MSSM to SM to be smaller than the inherent accuracy of
eq.(\ref{gaussian}).

When comparing MLLA predictions with experiments, the overall
normalization $\bar n_i$ (which depends on energy) is usually taken
from data. We cannot follow this approach here, since no data with $Q
\sim M_X$ are available. Moreover, usually MLLA predictions are
compared with inclusive spectra of all (charged) particles. We need
separate predictions for various kinds of hadrons, and are therefore
forced to make the assumption that all these FFs have the same
$x-$dependence at small $x$. This is perhaps not so unreasonable; we
saw above that the DGLAP evolution predicts such a universal
$x-$dependence at small $x$. We then match these analytic solutions
(\ref{gaussian}), (\ref{gauss_coef}) with the hadronic FFs $D_I^h$ we
obtained from DGLAP evolution and our input FFs at values $x_0^h$,
where for each hadron species $h$ the matching point $x_0^h$ and the
normalization $\bar n_h$ are chosen such that the FF and its first
derivative are continuous; we typically find $x_0 \sim 10^{-4}$. Note
that this matching no longer allows to respect energy conservation
exactly. However, since the MLLA solution begins to deviate from the
original FFs only at $x \sim 10^{-7}$, the additional ``energy
losses'' are negligible.

\begin{figure}
\input{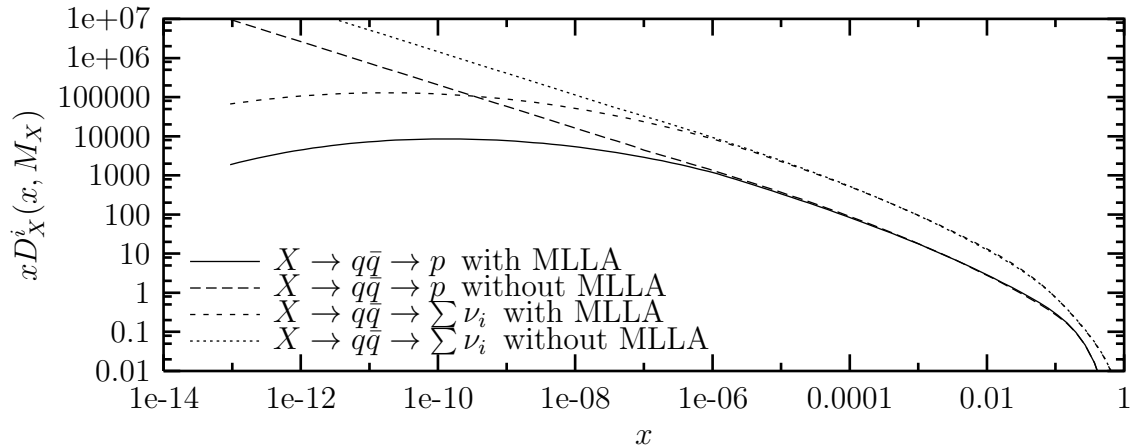}
\caption{Comparison between the MLLA solution and our results without
coherence effects, for the final proton and neutrino spectra. We
assume that $X$ undergoes two--body decay into $q_L \bar q_L$.}
\label{MLLA}
\end{figure}

Some results of our MLLA treatment are shown in Fig.~\ref{MLLA}. Here
the ``non--MLLA'' curves have been obtained by extrapolating our
numerical results described earlier, which extend ``only'' to $x =
10^{-7}$, by using simple power--law fits.  We see that at $x \sim
10^{-11}$ the FFs are suppressed by about two orders of magnitude, but
the effect diminishes quickly at larger values of $x$. Note that the
FFs into protons and into neutrinos have slightly different shapes in
the small$-x$ region. By assumption the FFs have the same shape for
all {\em hadrons}; however, in going from the spectrum of pions and
kaons to the neutrino spectrum, several additional convolutions are
required, which shift the peak of the distribution to even smaller
values of $x$. This figure also shows that the MLLA predictions
closely tracks the non--MLLA solution for $x$ values that are several
orders of magnitude smaller than the matching point $x_0$; this
illustrates the advantage of requiring both the FF and its first
derivative to be continuous at $x_0$.

\section{Summary and Conclusions}
\label{sec:conclusion}
\setcounter{footnote}{1}

In this article, we presented a detailed analysis of the decay of a
very massive $X$ particle, extending our earlier work
\cite{BarbotDrees:1}. In particular, we were able to improve the
accuracy of our code; as a result, we can now ensure energy
conservation to a numerical accuracy of better than 1\%, as compared
to up to several \% in ref.\cite{BarbotDrees:1}. Moreover, we showed
that the dependence of our results on the necessary extrapolation of
the measured fragmentation functions (FFs) towards small $x$ is
negligible. We also included leading higher--order QCD corrections at
very small $x$ using the MLLA approximation for taking into account
color coherence effects; this approximation is in good agreement with
data from particle colliders. These effects become significant for $x
\lsim 10^{-7}$, decreasing the predicted fluxes at $x \sim 10^{-11}$
by about two orders of magnitude.  

Furthermore, we showed that varying SUSY parameters can have some
impact on our results, affecting the shapes of the FFs at $x \geq
0.01$ and in some cases also the total multiplicity; however, the
dependence on the SUSY spectrum is much milder than the dependence on
the primary $X$ decay mode(s). Qualitatively the photon and LSP fluxes
are the most important ones at large $x$ if the primary is a strongly
interacting (s)particle; if the primary has only weak interactions,
the lepton fluxes can also be very large at large $x$. The proton flux
is always subdominant in this region. In contrast, the shapes of most
FFs at small $x$ can be predicted almost uniquely. This leads to the
following ordering of the fluxes at $x < 0.01$: the largest flux is of
muon neutrinos, followed by photons, $\nu_e$ and electrons, and
finally protons. The ratios of these fluxes become almost independent
of $x$ in this region, the proton flux being about a factor of five
smaller than the $\nu_\mu$ flux. On the other hand, the two smallest
fluxes at small $x$, of LSPs and finally $\nu_\tau$, do depend
sensitively on various currently unknown parameters. Generically they
rise less rapidly with decreasing $x$ than the other fluxes do;
already at $x \sim 10^{-3}$, the $\nu_\tau$ and LSP flux are usually
about one order of magnitude below the proton flux.

Finally, in the appendices we give additional details of our
description of the complete cascade. In particular, Appendix A
contains the first complete set of leading order splitting functions
for the MSSM, including all gauge as well as third generation Yukawa
interactions. A ``catalogue'' containing an almost complete set of FFs
for a given set of parameters is given in Appendix E.

This work presents the to date most accurate and complete description
of the spectra at source of stable particles resulting from the decay
of a superheavy $X$ particle. These spectra are needed for all
quantitative tests of the ``top--down'' explanation of the most
energetic cosmic ray events. Of course, in order to be able to compare
with fluxes measured on or near Earth, effects due to the propagation
through the galactic, and perhaps extragalactic, medium
\cite{reviewSigl} have to be included, which depend on the
distribution of $X$ particles throughout the Universe; we have made no
attempt to do this. On the other hand, our description of $X$ decays
is model--independent in the sense that it allows to incorporate any
primary $X$ decay mode. Indeed, it could with very little
modification also be used to describe the evolution of very energetic
jets produced through some other mechanism (e.g. the annihilation of
very massive stable particles), as long as the initial virtuality of
the produced particles is comparable to their energy.

Returning to the original problem of ultra--high energy cosmic rays
(UHECR), the biggest obstacle towards a test of generic top--down
models is the strong dependence of the predicted decay spectra on the
primary decay mode. Most previous investigations assumed that $X$
decays into a pair of quarks, but we are not aware of any compelling
argument why this should be the dominant decay mode. On the other hand,
data may already rule out some classes of top--down models. For
example, it seems likely that few, if any, UHECR are photons
\cite{nophot}. In the context of top--down models, this leaves protons
as only choice. Our results then seem to disfavor models where $X$
decays primarily into particles with only weak interactions, since
this implies a large ratio of the photon to proton flux at large
$x$. However, this argument may not apply if $M_X \gsim 10^{13}$ GeV,
since then all events seen so far are at $x \lsim 0.01$, where the
ratio of photon to proton fluxes is essentially independent of the
primary $X$ decay modes. Moreover, the photon flux may be diminished
more efficiently between source and detector than the proton
flux. Searches for very energetic neutrinos might therefore lead to
somewhat more robust tests of top--down models \cite{neutrinos,
bdhh1}; as noted earlier, the predicted neutrino flux should begin to
exceed the background from atmospheric neutrinos at very small values
of $x$. Nevertheless, the need to normalize the expected flux to the
observed flux of UHECR events, and hence to the proton and perhaps
photon flux at much larger $x$, re--introduces a large model
dependence even in this case \cite{bdhh1}. Moreover, other proposed
explanations of the UHECR also predict sizable neutrino fluxes at very
high energy, e.g. due to the GZK process itself. The failure to
observe such neutrinos could therefore exclude top--down models (given
sufficiently large detectors), but a positive signal may not be
sufficient to distinguish them from generic ``bottom--up''
models. This discrimination might be achieved by searching for the
predicted flux of very energetic LSPs, since the LSP flux in
bottom--up models is undetectably small; however, this test will
require very large detectors \cite{bdhh2}. We conclude that ultimately
the test of this idea will probably require a combined analysis of
different signals, at quite different energies and in different
detectors. We provide one of the tools needed to perform such an
analysis, since we are able to systematically study the fluxes of {\em
all} stable particles at source, and their correlations, for {\em all}
top--down models.

\section*{Acknowledgements}

We would like to thank Willy van Neerven for helpful comments on
higher order corrections. This work was supported in part by the
SFB375 of the Deutsche Forschungsgemeinschaft.

\setcounter{equation}{0}
\renewcommand{\theequation}{A.\arabic{equation}}

\section*{Appendix A: Splitting functions of the MSSM}
\label{subsec: SF}

The splitting function (SF) $P_{ji}(x)$ describes the radiation of
particle $j$ off particle $i$. Its $x-$dependence is determined by the
Lorentz structure of the corresponding vertex, while the normalization
also depends on the associated group [color and $SU(2)$] factors. If
there is no vertex relating these two particles the SF is simply 0. We
first list the functional forms we will need, together with the spins
of the particles involved in the branching process $i \rightarrow j +
k$ ($V$ for vector, $F$ for spin$-1/2$ fermion, $S$ for scalar):
\beqa
\label{e9}
&& (0)\;\;\; \delta(1-x)\,,\nonumber\\
&& (1)\; i=F, j=F, k=V:\ \frac{1 + x^2} { (1-x)_+ } \,, \nonumber \\
&& (2)\; i=F, j=V, k=F:\ \frac{1 + (1-x)^2} {x} \,, \nonumber \\
&& (3)\; i=F, j=S: k=F:\  x\,, \nonumber \\
&& (4)\; i=F, j=F, k=S:\ (1-x)\,,\nonumber\\
&& (5)\; i=S, j=F, k=F:\ 1\,,\nonumber\\
&& (6)\; i=S, j=V, k=S:\ \frac {2(1-x)} {x}\,,\nonumber\\
&& (7)\; i=S, j=S, k=V:\ \frac {2x} {(1-x)_+}\,,\nonumber\\
&& (8)\; i=V, j=F, k=F:\ (1-x)^2 + x^2\,,\nonumber\\
&& (9)\; i=V, j=V, k=V:\  2 \left[ \frac{1-x}{x} + x(1-x) +
\frac{x}{(1-x)_+} \right] \,,\nonumber\\
&& (10)\; i=V, j=S, k=S:\  2x(1-x)\,. 
\eeqa
For convenience, we also define $(1') = (1)+(0)$ and $(7')=(7)+(0)$.

\begin{center}
\begin{table}[ht]
\begin{center}
\begin{tabular}{|c||c|c|c|c|}\hline
$\alpha_S$ &\rule[-3mm]{0mm}{8mm}$ i=q $ & $\tilde{q}$ & $g$ &
$\tilde{g}$ 
\\ \hline
$j=q$ & \rule[-3mm]{0mm}{8mm} $\frac{4}{3}$ $(1^{'})$ & $\frac{4}{3}$
(5) & $\frac{N_q}{2}$ (8) & $\frac{N_q}{2}$ (4)
\\ \hline
$ \tilde{q}$ & \rule[-3mm]{0mm}{8mm} $\frac{4}{3}$ (3) & $\frac{4}{3}$
$(7^{'})$ & $\frac{N_q}{2}$ (10) & $\frac{N_q}{2}$ (3) 
\\ \hline
$g$ & \rule[-3mm]{0mm}{8mm} $\frac{4}{3}$ (2) & $\frac{4}{3}$ (6) & $ 3
\left[ (9) + \left( \frac{3}{2} - \frac{F}{6} \right) (0) \right] $ &
3 (2)
\\ \hline
$\tilde{g}$ & \rule[-3mm]{0mm}{8mm} $\frac{4}{3}$ (4) & $\frac{4}{3}$
(5) & 3 (8) & $ 3 \left[ (1) + \left( \frac{3}{2} - \frac{F}{6}
\right) (0) \right]$
\\ \hline
\end{tabular}
\caption{SUSY--QCD splitting functions $P_{ji}$, where $j$ and $i$
determine the row and column of the table, respectively. The
functional forms denoted by $(n), \ n=0, \dots, 10$ have been defined
in eq.(\ref{e9}), with $(1') = (1)+(0)$ and $(7') = (7) + (0)$. The
``multiplicity factors'' are: $N_{t_R} = N_{b_R} = 1$, $N_{t_L} =
N_{u_R} = N_{d_R} = 2$ and $N_{q_L} = 4$. In the MSSM phase, i.e. for
$Q > M_{\rm SUSY}$, the number of active flavors (quarks and squarks)
is $F=6$.}
\label{alphaS}
\end{center}
\end{table}
\end{center}

The 16 SFs of SUSY--QCD listed in table~\ref{alphaS} are derived from
\cite{Jones}; in eq.(\ref{e8}) they come with a factor of the strong
coupling $\alpha_S$. Note that in ref.\cite{Jones} the chirality index
$L,R$ was always summed over; e.g. $P_{tg} = P_{t_Lg} + P_{t_R g}$,
where $t_L$ now {\em only} describes the left--handed top quark (and
not the third generation quark doublet). Since these two terms are
equal, one has $P_{t_L g} = P_{t_R g} = P_{qg}/2$. On the other hand,
our ``(s)quark'' distributions always include anti(s)quarks. This
re--introduces a factor of 2, so that for us e.g. $P_{t_R g} = P_{qg}$
of \cite{Jones}. Additional factors arise for (s)quarks of the first
and second generation. As described in Sec.~2.2, we always average
over (s)quarks and anti(s)quarks with given hypercharge of the first
two generations. This implies $P_{u_R g} = P_{d_R g} = 2 P_{qg}$ and
$P_{q_L g} = 4 P_{qg}$, where the additional factor of two in the
second expression comes from summing over the $SU(2)$ index of the
doublet $q_L$. The same factors appear in SFs describing gluon to
squark splitting as well as gluino splitting into a squark and a
quark. A complete list of these factors $N_q$ is given in the table
caption.  On the other hand, in the absence of flavor--changing
interactions SFs involving quarks and squarks only always come with
factor 1 if the ``compound particles'' $u_R, q_L$ etc. are properly
normalized.

The SFs stemming from electroweak interactions have similar
structures; we just need to compute the correct group and multiplicity
factors. The results are listed in tables~\ref{g2} and \ref{gY}. In
these tables we list SFs including the appropriate multiplicity
factor; a single $SU(2)$ doublet {\em without} antiparticles would
have $N_f = 1/2$. Note that there is no difference between Higgs and
$SU(2)$ doublet lepton superfields as far as gauge interactions are
concerned. Finally, due to the absence of self--interactions of $U(1)$
gauge bosons, the splitting functions $P_{BB}$ and $P_{\tilde B \tilde
B}$ are pure delta--functions, with coefficients fixed by energy
conservation, eq.(\ref{sumrule}). In all three gauge interactions we
find that the coefficient of the $\delta-$function is the same in
$P_{ff}$ and $P_{\tilde f \tilde f}$ for any matter fermion $f$, and
also in $P_{VV}$ and $P_{\tilde V \tilde V}$ for a gauge boson $V$;
this latter coefficient is $-1/2$ times the coefficient in the
$\beta-$function of the corresponding gauge coupling.

\begin{center}
\begin{table}[h]
\begin{center}
\begin{tabular}{|c||c|c|c|c|}\hline
$g_2 = e/\sin
\theta_W$ &\rule[-3mm]{0mm}{8mm} $i=W$ & $\tilde{W}$ & $f_L$ & $\tilde
f_L$ 
\\ \hline
$j=W$ & \rule[-3mm]{0mm}{8mm} $2 \left[(9) + \left( \frac{3}{2} -
\frac{N_d}{8} \right) (0)\right] $ & 2 (2) & $\frac{3}{4}$ (2) &
$\frac{3}{4}$ (6) 
\\ \hline
$\tilde{W}$ & \rule[-3mm]{0mm}{8mm}2 (8) & 2 $\left[ (1) + \left(
\frac{3}{2} - \frac{N_d}{8} \right) \,(0) \right]$ & $ \frac{3}{4}$
(4) & $\frac{3}{4}$ (5) 
\\ \hline
$f_L$ & \rule[-3mm]{0mm}{8mm} $\frac{N_f}{2}$ (8) & $\frac{N_f}{2}$
(4) & $\frac{3}{4}$ $(1')$ & $\frac{3}{4}$ (5)
\\ \hline
$\tilde f_L$ & \rule[-3mm]{0mm}{8mm} $\frac{N_f}{2}$ (10) &
$\frac{N_f}{2}$ (3) & $\frac{3}{4}$ (3) & $\frac{3}{4}$ $(7^{'})$
\\ \hline
\end{tabular}
\caption{$SU(2)$ splitting functions $P_{ji}$, where particles $j$ and
$i$ are associated with the row and column, respectively. The
functional forms denoted by $(n), \ n=0, \dots, 10$ have been defined
in eq.(\ref{e9}), with $(1') = (1)+(0)$ and $(7') = (7) + (0)$. $N_d$
is the total number of $SU(2)$ doublets; in the MSSM, $N_d = 14$. $f$
stands for any matter or Higgs fermion, with $N_f$ being the number of
doublets (not counting anti--doublets) described by $f_L$ or $\tilde
f_L$. For our ``compound'' states, these are: $N_{q_L} = 6$, $N_{l_L}
= 2$, $N_{t_L} = 3$, $N_{\tau_L} = N_{H_1} = N_{H_2} = 1$.}
\label{g2}
\end{center}
\end{table}
\end{center}

Finally, Yukawa couplings only appear in $H f_L f_R, \ \tilde h \tilde
f_L f_R$ and $\tilde h f_L \tilde f_R$ vertices. We therefore only
need functional forms (3), (4) and (5) from eq.(\ref{e9}). The
coefficients can be determined from the analogous terms due to
$U(1)_Y$ interactions by replacing $ (g_Y Y_f)^2$ by $\lambda_f^2/2$,
where the extra factor of $1/2$ corrects for the factor $\sqrt{2}$
appearing in front of gaugino--fermion--sfermion

\clearpage

\begin{center}
\begin{table}
\begin{center}
\begin{tabular}{|c||c|c|c|c|} \hline
$g_Y = e/\cos \theta_W$ & \rule[-3mm]{0mm}{8mm} $i=B$ & $\tilde{B}$ &
$f$ & $\tilde f$ 
\\ \hline
$j=B$ & \rule[-3mm]{0mm}{8mm} $-\frac{1}{2} \sum_f Y_f^2$ (0) & 0 &
$Y_f^2$ (2) & $Y_f^2$ (6)
\\ \hline
$\tilde{B}$ & \rule[-3mm]{0mm}{8mm} 0 & $-\frac{1}{2} \sum_f Y_f^2$ 
(0) & $Y_f^2$ (4) & $Y_f^2$ (5) 
\\ \hline
$f$ & \rule[-3mm]{0mm}{8mm} $ n_f Y_f^2 $ (8) & $ n_f Y_f^2$ (4) &
$Y_f^2$ (1') & $Y_f^2$ (5) 
\\ \hline
$\tilde f$ & \rule[-3mm]{0mm}{8mm}$ n_f Y_f^2$ (10) & $ n_f Y_f^2$ (3)
& $Y_f^2$ (3) & $Y_f^2$ (7') 
\\ \hline
\end{tabular}
\caption{$U(1)_Y$ splitting functions $P_{ji}$, where particles $j$
and $i$ are associated with the row and column, respectively. The
functional forms denoted by $(n), \ n=0, \dots, 10$ have been defined
in eq.(\ref{e9}), with $(1') = (1)+(0)$ and $(7') = (7) + (0)$. The
sum of squared hypercharges of all particles $\sum_f
Y_f^2 = 11$ in the MSSM. $f$ stands for any matter or Higgs fermion
with hypercharge $Y_f$, while $n_f$ is the number of degrees of
freedom (not counting anti--particles) described by $f$ or $\tilde
f$. For our ``compound'' states, these are: $Y_{q_L}^2 = 1/36, n_{q_L}
= 12$; $Y^2_{u_R} = 4/9, n_{u_R} = 6$; $Y^2_{d_R} = 1/9, n_{d_R} =
6$; $Y^2_{l_L} = 1/4, n_{l_L} = 4$; $Y^2_{l_R} = 1, n_{l_R} = 2$;
$Y^2_{t_L} = 1/36, n_{t_L} = 6$; $Y^2_{t_R} = 4/9, n_{t_R} = 3$;
$Y^2_{b_R} = 1/9, n_{b_R} = 3$; $Y^2_{\tau_L} = Y^2_{H_1} = Y^2_{H_2}
= 1/4, n_{\tau_L} = n_{H_1} = n_{H_2} = 2$; $Y^2_{\tau_R} = 1,
n_{\tau_R} = 1$.}
\label{gY}
\end{center}
\end{table}
\end{center}
\begin{center}
\begin{table}
\begin{center}
\begin{tabular}{|c||c|c|c|c|c|c|}\hline
$\lambda_f$ & $i=H$ & $\tilde H$ & $f_L$ & $\tilde f_L$ & $f_R$ &
$\tilde f_R$
\\ \hline
$j=H$ & \rule[-3mm]{0mm}{8mm} $-\frac{N_c}{2}$ (0) & 0 & $\frac{1}{2}$
(3) & 0 & (3) & 0 \\ \hline
$\tilde H$ & \rule[-3mm]{0mm}{8mm} 0 &
$-\frac{N_c}{2}$ (0) & $\frac{1}{2}$ (4) & $\frac{1}{2}$ (5) & (4) & (5)
\\ \hline
$f_L$ & \rule[-3mm]{0mm}{8mm} $\frac{N_c}{2}$ (5) & $\frac{N_c}{2}$
(4) & $-\frac{1}{2}$ (0) & 0 & (4) & (5)
\\ \hline
$\tilde f_L$ & \rule[-3mm]{0mm}{8mm} 0 & $\frac{N_c}{2}$ (3) & 0 &
$-\frac{1}{2}$ (0) & (3) & 0
\\ \hline
$f_R$ & \rule[-3mm]{0mm}{8mm} $\frac{N_c}{2}$ (5) & $\frac{N_c}{2}$
(4) & $\frac{1}{2}$ (4) & $\frac{1}{2}$ (5) & $-1$ (0) & 0
\\ \hline
$\tilde f_R$ & \rule[-3mm]{0mm}{8mm} 0 & $\frac{N_c}{2}$ (3) 
& $\frac{1}{2}$ (3) & 0 & 0 & $-1$ (0)
\\ \hline
\end{tabular}
\caption{Splitting functions $P_{ji}$ originating from Yukawa
interactions, where particles $j$ and $i$ are associated with the row
and column, respectively. The functional forms denoted by $(n), \ n=0,
3, 4, 5$ have been defined in eq.(\ref{e9}). Since we only include
Yukawa interactions for the third generation, we only have to consider
three cases. For the top Yukawa coupling, $f_L = t_L, \, f_R = t_R, \,
H = H_2$ and number of colors $N_c = 3$; for the bottom Yukawa
coupling, $f_L = t_L, \, f_R = b_R, \, H = H_1$ and $N_c = 3$;
finally, for the tau Yukawa coupling, $f_L = \tau_L, \, f_R = \tau_R,
\ H = H_1$ and $N_c = 1$.}
\label{Yuk}
\end{center}
\end{table}
\end{center}

\clearpage

\noindent
 vertices in the
supersymmetric Lagrangian.  Since Yukawa interactions couple matter
fields with different chiral indices all diagonal SFs due to Yukawa
couplings are pure $\delta-$functions, the coefficients again being
determined by energy conservation; as before, we find equal
coefficients for diagonal SFs of a particle and its superpartner. The
resulting SFs are listed in table~\ref{Yuk}. As usual, these SFs are
multiplied with $\alpha_f/(2\pi) \equiv \lambda_f^2/(8 \pi^2)$ in the
DGLAP equations.  The three interactions we consider, involving the
top, bottom and tau Yukawa couplings, can all be treated using
table~\ref{Yuk}, by identifying the matter and Higgs fields
appropriately and using the correct color factors, as explained in the
caption. The $SU(2)$ factors, which lead to the factor of 2 difference
between SFs describing radiation off $SU(2)$ singlet or doublet
(s)fermions, are the same in all three cases.\footnote{Strictly
speaking, $H_1$ can only split into $\tau_R$ and $\overline{\tau_L}$,
not into $\tau_L$ and $\overline{\tau_R}$, while the antiparticle
$H_1^*$ can only split into $\tau_L$ and $\overline{\tau_R}$;
analogous remarks hold for the other Yukawa--induced branching
processes. However, this distinction plays no role for us, since we
always average or sum over particle and antiparticle.}

\setcounter{equation}{0}
\renewcommand{\theequation}{B.\arabic{equation}}

\section*{Appendix B: Unitary transformations between current and mass
eigenstates in the MSSM} 
\label{sec: Transfo}

In this Appendix we describe the unitary transformations occurring
during the SUSY and $SU(2) \otimes U(1)$ breaking, where the quarks,
leptons, weak gauge bosons and Higgs bosons as well as all
superparticles acquire their masses \cite{Martin}. The superscript $b$
denotes the mass eigenstates of the broken theory. The fields in the
unbroken theory are the same as those described in Sec.~2.2. For
example, $q_L$ stands for all left--handed quarks and antiquarks of
the two first generations, i.e. the $SU(2)$ doublets $(u_L,d_L)$,
$(c_L,s_L)$ and their antiparticles $(\bar{u}_L,\bar{d}_L)$,
$(\bar{c}_L, \bar{s}_L)$, and thus describes eight degrees of freedom
(times three, if color is counted separately). Similarly, $l_L$ stands
for both $SU(2)$ doublets $(e_L,\nu_e)$ and $(\mu_L,\nu_\mu)$ and
their antiparticles $(\bar{e}_L,\bar{\nu}_e)$ and
$(\bar{\mu}_L,\bar{\nu}_\mu)$. On the other hand, $u^b$ only describes
$u-$quarks and their antiparticles, but includes both chirality
states, and thus describes four degrees of freedom (not counting
color). Recall that the transformation between mass and current
eigenstates in eq.(\ref{trafo}) only affects the {\em upper} index of
the (generalized) FFs. In the given context $q_L$ therefore stands for
the sum, not the average, of its ``constituent fields'', as discussed
in Sec.~2.2. Recall finally that massive gauge bosons ``eat''
Goldstone modes from the Higgs sector. These considerations lead to
the following transformations for SM fields and Higgs bosons:
\beqa \label{smtrafo}
u^b &=& c^b = \frac{1}{4}\, q_L + \frac{1}{2}\, u_R\,,\nonumber\\
d^b &=& s^b = \frac{1}{4}\, q_L + \frac{1}{2}\, d_R\,,\nonumber\\
b^b &=& \frac{1}{2}\, t_L + b_R\,,\nonumber\\
t^b &=& \frac{1}{2}\, t_L + t_R\,,\nonumber\\\nonumber\\
e^b &=& \mu^b = \frac{1}{4}\,l_L + \frac{1}{2}\,e_R\,,\nonumber\\
\tau^b &=& \frac{1}{2}\,\tau_L + \tau_R\,,\nonumber\\
\nu^b_e &=& \nu^b_{\mu} = \frac{1}{4}\,l_L\,,\nonumber\\
\nu^b_{\tau} &=& \frac{1}{2}\,\tau_L\,,\nonumber\\
g^b &=& g\,,\nonumber\\
W^b &:=& W^+ + W^- = 2 \left( \frac{1}{3} W + \cos^2{\beta} \frac {H_1}{4} +
\sin^2{\beta} \frac {H_2}{4} \right) \, , \nonumber \\
Z^b &=& \sin^2(\theta_W) \,B + \cos^2(\theta_W) \, \frac {W}{3} +
\cos^2{\beta} \frac {H_1}{4} + \sin^2{\beta} \frac {H_2}{4} \, ,
\nonumber \\
\gamma^b &=& \cos^2(\theta_W)\,B + \sin^2(\theta_W)\, \frac{W}{3} \, ,
\nonumber \\
h^{0\,b} &=& \sin^2{\alpha} \, \frac {H_1}{4} + \cos^2{\alpha} \,
\frac {H_2} {4} \, , \nonumber \\
H^{0\,b} &=& \cos^2{\alpha} \, \frac {H_1}{4} + \sin^2{\alpha} \,
\frac {H_2} {4} \, , \nonumber \\
A^{0\,b} &=& \sin^2{\beta} \, \frac {H_1}{4} + \cos^2{\beta} \, \frac
{H_2}{4} \, , \nonumber \\
H^b &:=& H^+ + H^- = 2 \left( \sin^2{\beta} \, \frac {H_1}{4} +
\cos^2{\beta} \, \frac {H_2}{4} \right) \, . 
\eeqa

All superparticles also acquire masses at this stage, and the
particles with identical quantum numbers mix together to give the
mass eigenstates:
\beqa \label{susytrafo}
\tilde{q}_{L/R}^b &=& \tilde{q}_{L/R} \, \, {\rm for} \, q = u,d,s,c
\, , \nonumber \\
\tilde{b}_1^b &=& \frac{1}{2} \cos^2(\theta_b) \, \tilde{t}_L +
\sin^2(\theta_b) \, \tilde{b}_R \, , \nonumber \\
\tilde{t}_1^b &=& \frac{1}{2} \cos^2(\theta_t) \, \tilde{t}_L +
\sin^2(\theta_t) \, \tilde{t}_R \, , \nonumber \\
\tilde{b}_2^b &=& \frac{1}{2} \sin^2(\theta_b) \, \tilde{t}_L +
\cos^2(\theta_b) \, \tilde{b}_R \, , \nonumber \\
\tilde{t}_2^b &=& \frac{1}{2} \sin^2(\theta_t) \, \tilde{t}_L +
\cos^2(\theta_t) \, \tilde{t}_R \, , \nonumber \\
\tilde{e}_L^b &=& \tilde{\mu}_L^b = \frac{1}{4} \, \tilde{l}_L \, ,
\nonumber \\
\tilde{e}_R^b &=& \tilde{\mu}_R^b = \frac{1}{2} \, \tilde{e}_R \, ,
\nonumber \\
\tilde{\tau}_1^b &=& \frac{1}{2} \, \cos^2(\theta_\tau) \, \tilde{\tau}_L +
\sin^2(\theta_\tau) \, \tilde{\tau}_R \, , \nonumber \\
\tilde{\tau}_2^b &=& \frac{1}{2} \, \sin^2(\theta_\tau) \, \tilde{\tau}_L +
\cos^2(\theta_\tau) \, \tilde{\tau}_R \, , \nonumber \\
\tilde{\nu}_e^b &=& \tilde{\nu}_\mu^b = \frac{1}{4} \, \tilde{l}_L \,
, \nonumber \\
\tilde{\nu}_\tau^b &=& \frac{1}{2} \, \tilde{\tau}_L \, , \nonumber \\
\tilde{g}^b &=& \tilde{g} \, , \nonumber \\
\tilde{\chi}_1^b &:=& \tilde{\chi}_1^+ + \tilde{\chi}_1^- =
\left[ \sin^2(\gamma_R) + \sin^2(\gamma_L) \right] \, \frac
{\tilde{W}}{3} + \cos^2(\gamma_R) \, \frac{\tilde{H}_2}{2} +
\cos^2(\gamma_L) \, \frac{\tilde{H}_1}{2} \, , \nonumber \\
\tilde{\chi}_2^b &:=& \tilde{\chi}_2^+ + \tilde{\chi}_2^- =
\left[ \cos^2(\gamma_R) + \cos^2(\gamma_L) \right] \, \frac
{\tilde{W}}{3} + \sin^2(\gamma_R) \, \frac {\tilde{H}_2}{2} +
\sin^2(\gamma_L) \, \frac {\tilde{H}_1}{2} \, , \nonumber \\
\tilde{\chi}_i^{0b} &=& 
\left| v_1^{(i)} \right|^2 \frac {\tilde H_1}{2}
+ \left| v_2^{(i)} \right|^2 \frac {\tilde H_2}{2}
+ \left| v_3^{(i)} \right|^2 \frac {\tilde W}{3}
+ \left| v_4^{(i)} \right|^2 {\tilde B}.
\eeqa
Here we have largely followed the notation of ISASUSY \cite{Isasusy}.
However, we have used the more common symbol $\tilde \chi$ for
charginos and neutralinos; in ISASUSY notation, $\tilde \chi_1^b =
\widetilde W_-, \ \tilde \chi_2^b = \widetilde W_+$, and $\tilde
\chi_i^{0b} = \widetilde Z_i$. The mixing angles $\alpha$ (in the
Higgs sector), $\theta_b, \, \theta_t, \, \theta_\tau$ (in the
sfermion sector), $\gamma_L, \, \gamma_R$ (in the chargino sector) as
well as the $v_i^{(j)}$ (in the neutralino sector) have been computed
numerically using ISASUSY.

\setcounter{equation}{0}
\setcounter{footnote}{0}
\renewcommand{\theequation}{C.\arabic{equation}}

\section*{Appendix C: Two-- and three--body decay spectra}
\label{sec:Decays}

\subsection*{Generalities}

We want to define the decay functions (DFs) $\tilde{P}_{sS}$
describing two-- or three--body decay $S \rightarrow s$, see
eq.(\ref{decay}). These DFs can be obtained directly from the decay
spectrum of $S$ in the ultra--relativistic limit, where the energy
$E_S$ is much larger than the mass $M$ of $S$:
\beq \label{decff}
\tilde P_{sS}(z) = \frac {1} {\Gamma} \frac {d \Gamma(E_S)} {d z},
\eeq
where $z = E_s / E_S$. This spectrum can e.g. be evaluated by first
computing the double differential decay distribution $d^2 \Gamma / (d
E_s^* d \cos \theta^*)$ in the {\em rest frame} of $S$, then boosting
the four--momentum of $s$ with boost factor $\gamma = E_S / M$ at
angle $\theta^*$ relative to $\vec{p}_s$, and finally integrating over
$\cos \theta^*$ subject to the constraint that the boosted energy of
$s$ equals $E_s$. Note that eq.(\ref{decff}) implies $\int_0^1 \tilde
P_{sS}(z) dz = 1$; if $S-$decays produce $N$ identical particles $s$,
the corresponding $\tilde P_{sS}$ would thus have to be multiplied
with an extra factor of $N$, in order to correctly reproduce the total
multiplicity in the final state. Finally, momentum conservation
implies $\sum_s \int_0^1 z \tilde P_{sS}(z) = 1$.

In case of two--body decays $S \rightarrow i + j$ the energy $E_s^*$
in the rest frame of $S$ is fixed completely by the kinematics. The
boost and integration over $\cos \theta^*$ then leads to a flat
decay function:
\beq
\tilde{P}^{(2)}_{iS} (z) =
\left\{ \left[ 1 - \left( \frac {m_1+m_2} {M } \right)^2 \right]
\left[ 1 - \left( \frac {m_1-m_2} {M} \right)^2 \right]
\right\}^{-\frac{1}{2}} \Theta(z - z^{(i)}_{-})\, \Theta(z^{(i)}_{+}-z)
\eeq
for the decay product $i$ with $i = 1$ or $2$. The kinematic minimum
and maximum $z^{(i)}_{\pm}$ of $z$ are given by:
\beq \label{zlim}
z^{(i)}_{\pm} = \frac{1}{2} \left( 1 + \frac {m_i^2 - m_j^2} {M^2} \pm
\sqrt{ \left[ 1 - \left( \frac {m_1+m_2}{M} \right)^2 \right] \left[ 1
- \left( \frac {m_1-m_2}{M} \right)^2 \right]} \right).
\eeq
For example, for $m_1 \rightarrow M, \ m_2 \rightarrow 0$,
eq.(\ref{zlim}) implies $z^{(1)}_\pm \rightarrow 1, \ z^{(2)}_\pm
\rightarrow 0$, i.e. the entire energy of $S$ goes into the massive
decay product. In contrast, for $m_1 = m_2$, the energy of $S$ will on
average be shared equally between the two decay products; if $m_1 =
m_2 \rightarrow 0$, the $z^{(i)}$ can lie anywhere between zero and
one. Since $E_s^*$ is fixed in this case, our treatment of two--body
decays is exact up to possible polarization effects; we do not expect
these effects to be very important, except perhaps in case of $\tau$
decays (which, however, usually do not contribute very much to the
final spectra of stable particles).

Three--body decays lead to a nontrivial distribution of the energy of
the decay products already in the rest frame of $S$. For simplicity we
assume that at most one of the three decay products is massive; this
should be a safe approximation, except for $b \rightarrow c \tau
\nu_\tau$ decays, which have a rather small branching ratio. We then
need separate DFs for the massive and massless decay products. For the
massive decay product, with mass $m$, we find
\beq
\tilde{P}^{(3)}_{sS}(z) = N_3 \left[ 1 - z + \frac{m^2}{M^2} \left(
1 - \frac{1}{z} \right) \right]
\eeq
where $z \in [\frac{m^2}{M^2},1]$ and the normalization factor is
given by:
\beq \label{n3}
N_3 = \left[ \frac{1}{2} \left( 1 - \frac{m^4}{M^4} \right) +
\frac{m^2}{M^2} \log \frac{m^2}{M^2} \right]^{-1}. 
\eeq
If on the contrary, $s$ is one of the massless decay products, we find:
\beq
\tilde{P}^{(3)}_{sS}(z) = N_3 \left [ 1 - z - \frac{m^2}{M^2} \left( 1 +
\log \frac{M^2}{m^2} + \log(1-z) \right) \right],
\eeq
where now $z \in [0,1-\frac{m^2}{M^2}]$; the normalization factor
$N_3$ has been given in eq.(\ref{n3}).

Our treatment of three--body decays is not exact, since it ignores
dynamical effects (described by the invariant Feynman amplitude) on
the decay spectrum in the $S$ rest frame.\footnote{The calculation of
the corresponding branching ratio in ISASUSY does include these
dynamical effects; in other cases the required branching ratio can be
taken directly from experiment, e.g. for $\tau$ decays.} However,
treating these effects properly is quite nontrivial, since it would
force us to introduce many different three--body decay functions. Note
in particular that massive superparticles (charginos and neutralinos)
do generally not decay via $V-A$ interactions, unlike the $b$ and $c$
quarks and heavy $\mu$ and $\tau$ leptons in the SM. Moreover, the
Feynman amplitudes in many cases depend nontrivially on the
polarization of the decaying particle; this could only be described at
the cost of introducing many additional generalized fragmentation
functions, since we would have to keep track of left-- and
right--handed particles separately. However, experience from hadron
collider physics teaches us that including the exact decay matrix
elements is usually not very important if one is only interested in
single--particle inclusive spectra. We expect this to be true in our
case as well, since the convolution with parton distribution functions
necessary at hadron colliders is reminiscent of the convolution with
generalized FFs in our case. We finally note that longer decay chains
involving two-- and three--body decays can be treated by simply
convoluting appropriate factors of $\tilde P^{(2)}_{sS}$ and $\tilde
P^{(3)}_{sS}$. 

\subsection*{Treatment of heavy quark decays}

The top quark being very heavy ($m_t \sim 175$ GeV $\gg m_{\rm had}
\sim 1 $ GeV), it decays before hadronizing, and can thus be included
in the decay cascade at scale $M_{\rm SUSY}$. On the other hand, the
hadronization of the $b$ and $c$ quarks has to be treated with some
care. The ``input'' fragmentation functions we used \cite{Poetter}
already include the final hadrons (nucleons, kaons and pions) produced
at the end of the decay cascade of $c-$ and $b-$flavored
hadrons. However, they do not include the leptons arising from this
cascade, which are not negligible. We therefore implemented a special
treatment for this component, using the empirical FFs proposed by
Peterson et al. \cite{Peterson} for heavy quarks as a basis for the
fragmentation of $c-$ and $b-$hadrons. To that end, we used two
``generic'' particles, a $c-$ and a $b-$hadron, with respective
average masses $\overline{m}_c = 2.1$ GeV and $\overline{m}_b = 5.3$
GeV; we also had to renormalize the complete set of FFs for $b$'s and
$c$'s. The scheme can be described by Fig.~12. Here, $B_l(b)$ and
$B_l(c)$ describe the branching ratio of the semi--leptonic decay
modes of $b-$ and $c-$flavored hadrons, respectively [summed over all
accessible pairs ($l$,$\nu_l$)].

\vspace*{35mm}
\begin{center} 
\begin{figure}[ht] \label{hqdec}
\begin{picture}(-400,50)(0,-120)
\SetOffset(440,0)
\label{heavy_decays}
\SetPFont{Helvetica}{24}
\Text(-405,0)[]{$b$}
\ArrowLine(-395,0)(-330,0) 
\Text(-360,-10)[]{Peterson} 
\Text(-300,0)[]{$b-$hadron}
\ArrowLine(-265,0)(-165,0)
\Text(-220,10)[]{semi-leptonic}
\Text(-220,-10)[]{decay $B_l(b)$}
\Text(-120,0)[]{$c-$hadron   $l$   $\nu_l$}
\Line(-130,-10)(-130,-30)
\ArrowLine(-130,-30)(-40,-30)
\Text(-85,-20)[]{semi-leptonic}
\Text(-85,-40)[]{decay $B_l(c)$}
\Text(-5,-30)[]{$s-$hadron   $l$   $\nu_l$}
\Text(-405,-80)[]{$c$}
\ArrowLine(-395,-80)(-330,-80) 
\Text(-360,-90)[]{Peterson} 
\Text(-300,-80)[]{$c-$hadron}
\ArrowLine(-265,-80)(-165,-80)
\Text(-220,-70)[]{semi-leptonic}
\Text(-220,-90)[]{decay $B_l(c)$}
\Text(-120,-80)[]{$s-$hadron   $l$   $\nu_l$}
\end{picture} 
\caption{Schematic hadronization and decay cascade for heavy
quarks $c$ and $b$. The ``$s-$hadrons'', mainly kaons, are already included
in the FFs given in \cite{Poetter}.}
\end{figure}
\end{center}
\vspace*{5mm}

As mentioned earlier, the leptonic $b$ and $c$ decay products have to
be included in the normalization of the FFs $D_b^h$ and $D_c^h$. To
that end, we introduce $R_c$ and $R_s$, the energy carried by the $c$
and $s$ quark in semi--leptonic $b-$ and $c-$ decays, respectively, as
well as $x_B$ and $x_D$, the energy fraction of the $b$ ($c$) quark
carried by the $b-$flavored ($c-$flavored) hadron. The latter are
given by
\beq
x_{B,D} = \int_0^1 z D_{\rm Pet}^{b,c}(z) \, dz \, ,
\eeq

where $D_{\rm Pet}^{b,c}$ is the Peterson FF \cite{Peterson} for $b$
and $c$ quarks, respectively; we took $\epsilon_c = 0.15, \,
\epsilon_b = 0.015$ for the single free parameter in these FFs. We
compute $R_c$ and $R_s$ from pure phase space, i.e.  we again ignore
possible effects of the Feynman amplitudes on the three--body decay
distributions. This gives:
\beqa \label{eqR}
R_c &=& \frac {1} {\overline{m}_b \Gamma(b \rightarrow c l \nu) }
\int_{\overline{m}_c}^{E_c,{\rm max}} d E_c \, E_c \, \frac {d
\Gamma(b \rightarrow c l \nu)} {d E_c} \nonumber \\
&=& \frac {(1 - r)^3} {3 (1 - r^2) - 6 r \log r} \, ,
\eeqa
where $r = \overline{m}_c^2 / \overline{m}_b^2 = 0.157$ for our choice
of average $b-$ and $c-$hadron masses; note that $R_c \rightarrow 1/3
\ (1)$ for $r \rightarrow 0 \ (1)$. Eq.(\ref{eqR}) can also be used
for the computation of $R_s$, with $\overline{m}_c \rightarrow
\overline{m}_s \simeq 0.5$ GeV, $\overline{m}_b \rightarrow
\overline{m}_c$. The FFs of \cite{Poetter} only include the hadrons
produced in the fragmentation and decays of the $c-$ and
$b-$quarks. Their normalization, which we need to know for the
necessary extrapolation of these FFs towards small $x$ as discussed in
Appendix D, is therefore given by
\beqa 
\frac{1}{x_B} \sum_h \int_0^1 zD_b^h(z)\,dz &=&  \frac{1-x_B}{x_B}
\nonumber \\
&+& [1 - B_l(b)] \cdot [1 - B_l(c)]  \nonumber \\ 
&+& \, B_l(b) \cdot [1 - B_l(c)] \cdot R_c  \nonumber \\
&+& \, [1 - B_l(b)] \cdot B_l(c) \cdot ( R_c\cdot R_s + 1 - R_c )
\nonumber \\ 
&+& \, B_l(b) \cdot B_l(c) \cdot R_c \cdot R_s  \nonumber \\
&=& \frac{1}{x_B} - B_l(b) \cdot (1-R_c) - R_c \cdot B_l(c) \cdot (1-R_s).  
\eeqa
The right hand side can be understood as follows: the first line
describes the contribution of the light hadrons produced when the
$b-$quark hadronizes into a $b-$flavored hadron; the second line
describes purely hadronic decays; the third line describes leptonic
primary $b$ decays followed by hadronic $c$ decays (in this case only
the fraction $R_c$ of the $b-$hadrons energy goes into hadrons); the
fourth line describes the hadronic energy fraction in the case of a
hadronic primary $b$ decay followed by leptonic $c-$decays; finally,
the fifth line describes the hadronic energy fraction after leptonic
decays in both the primary and secondary decays. The same holds for
$c-$hadron decays, up to the simplifying fact that ``$s-$hadrons'' are
already fully included into the FFs of \cite{Poetter}. We get: 
\beq
\frac{1}{x_D} \sum_h \int_0^1 zD_c^h(z)\,dz =  \frac{1}{x_D} - B_l(c)
+ B_l(c) \cdot R_s.  \eeq

\setcounter{equation}{0}
\renewcommand{\theequation}{D.\arabic{equation}}

\section*{Appendix D: Parameterization of the input fragmentation
  functions}
\label{sec: FFs}

Here we give the input fragmentation functions (FFs) we used to
describe the hadronization of quarks and gluons, taken from
\cite{Poetter}, and the parameters of the extrapolation we made in the
small $x$ region. 

The functions taken from ref~\cite{Poetter} are given with the
functional form $N x^\alpha (1-x)^\beta$ ; there are given in
table~\ref{Poetter}, at the scale where quarks and hadrons hadronize,
i.e. $Q_0 = max(m_q,Q_{had})$. We used the NLO results, excepted for the
s quark, for which we had to use the LO ones, because the NLO form
didn't allow us to impose energy conservation and continuity at low x.

As we showed in Sec.~\ref{subsec:non_pert}, the
final result at low $x$ depends very little on the chosen power law in
our parameterization
\beq \label{fpara}
f(x) = ax^{-\alpha'} + b\log{x} + c \, ,\ a > 0,
\eeq
once energy conservation has been imposed. Here we therefore only
give results for a parameterization where $\alpha'$ is taken to be
1. That is, we assume that the multiplicity due to non--perturbative
effects gets the same contribution for each decade of energy, if the
hadron's energy is small compared to that of the initial parton.

\begin{center}
\begin{table}
\begin{center}
\begin{tabular}{|c||c||c|c|c||c|c|}\hline
&&&&&&\\
$D_p^h(x,Q_0^2)$ & $p$ & $n$ & $\pi^\pm$ & $\pi^0$ & $K^\pm$ & $K^0$ \\ 
$= N x^\alpha (1-x)^\beta$&&&&&&\\ \hline 
$u$ & $N = 1.26$&0.63&0.448&0.224&0.178&4.96\\ 
& $\alpha = 0.0712$&0.0712&-1.48&-1.48&-0.537&0.0556\\
& $\beta = 4.13$&4.13&0.913&0.913&0.759&2.8\\ \hline  
$d$ & 0.63&1.26&0.448&0.224&4.96&0.178\\
&0.0712&0.0712&-1.48&-1.48&0.0556&-0.537\\
&4.13&4.13&0.913&0.913&2.8&0.759\\ \hline
$s$ & 4.08&4.08&22.3&11.15&0.259&0.259\\
&-0.0974&-0.0974&0.127&0.127&-0.619&-0.619\\
&4.99&4.99&6.14&6.14&0.859&0.859\\ \hline  
$c$ & 0.0825&0.0825&6.17&3.085&4.26&4.26\\
&-1.61&-1.61&-0.536&-0.536&-0.241&-0.241\\
&2.01&2.01&5.6&5.6&4.21&4.21\\ \hline  
$b$ & 24.3&24.3&0.259&0.1295&1.32&1.32\\
&0.579&0.579&-1.99&-1.99&-0.884&-0.884\\
&12.1&12.1&3.53&3.53&6.15&6.15\\ \hline  
$g$ & 1.56&1.56&3.73&1.865&0.231&0.231\\
&0.0157&0.0157&-0.742&-0.742&-1.36&-1.36\\
&3.58&3.58&2.33&2.33&1.8&1.8\\ \hline  
\end{tabular}
\caption{Input fragmentation
functions at small $x$, with functional form $N x^\alpha
(1-x)^\beta$, taken from \cite{Poetter} at $Q_0 = max(m_q,Q_{had})$.
We took their NLO results for u,d,c,b and g, but the LO result for
the s quark. See the text for further details.}
\label{Poetter}
\end{center}
\end{table}
\end{center}

In order to obtain a unique solution with the only two constraints at
our disposal (energy conservation and continuity of the FFs), we
imposed these constraints on the sum of the FFs $\sum_h D_i^h(x,Q^2)$,
where $i$ is a given initial parton, and $h$ runs over the final
hadrons. Of course, energy will be conserved independently for each
initial parton $i$. For each $i$ we define a cut--off $x_0^i$ which
defines the transition between the functions given in \cite{Poetter}
and our extrapolation. The $x_0^i$ have to be chosen such that the
equations of energy conservation admit a solution; a necessary (but
generally not sufficient) requirement is that the integral over the
original FFs satisfy $\sum_h \int_{x_0}^1 dz z \, D_i^h(z) < 1$. Our
requirement of continuity at $x_0$ implies that the final results depend
very little on the precise values of the $x_0^i$. For simplicity we
assume that all the $D_i^h(x,Q^2_{\rm had})$ for a given $i$ have the
same shape at small $x$; recall that purely perturbative effects
ensure that this is true after DGLAP evolution, which anyway greatly
reduces the sensitivity to the input. The normalizations for the
various hadrons can then be read off directly from the results of
ref.\cite{Poetter}, once $x_0$ has been determined. The results are
presented in table.~\ref{extrapolation}, which lists the cut-off
$x_0^i$, the coefficients $a, b, c$ of the functional form
(\ref{fpara}) describing the {\em sum} $\sum_h D_i^h$ for fixed parton
$i$, and the normalization coefficients\footnote{At first sight the
relative ordering of the $N_u^K, \, N_d^K$ factors may seem
counter--intuitive. Indeed, a $u-$quark should more readily fragment
into a charged Kaon than into a neutral one, whereas the opposite
behavior is expected for $d-$quarks. Recall, however, that here we are
only interested in the behavior at small $x$. In this case
ref.\cite{Poetter} finds the opposite behavior as at large $x$,
i.e. $u-$quarks indeed seem to be more likely to produce a {\em soft}
neutral kaon than a {\em soft} charged kaon.} $N_i^h$, so that $D_i^h
= N_i^h \sum_h D_i^h$; of course, $\sum_h N_i^h = 1 \ \forall i$.

\begin{center}
\begin{table}
\begin{center}
\begin{tabular}{|c||c||c|c|c||c|c|c|c|c|c|}\hline
initial parton & $x_0$ & $a$ & $b$ & $c$ & $p$ & $n$ & $\pi^\pm$ &
$\pi^0$ & $K^\pm$ & $K^0$ \\ \hline 
$u$ & 0.27 & 4.06 & -9.74 & -14.40 & 0.05 & 0.025 & 0.38 & 0.19 & 0.05 &
0.31 \\ \hline
$d$ & 0.27 & 4.06 & -9.74 & -14.40 & 0.025 & 0.05 & 0.38 & 0.19 & 0.31 &
0.05 \\ \hline
$s$ & 0.20 & 5.74 & -18.47 & -31.42 & 0.14 & 0.14 & 0.41 & 0.21 & 0.05 &
0.05 \\ \hline
$c$ & 0.27 & 4.06 & -4.16 & -6.24 & 0.05 & 0.05 & 0.30 & 0.15 & 0.22 &
0.22 \\ \hline
$b$ & 0.20 & 5.74 & -27.81 & -49.27 & 0.08 & 0.08 & 0.35 & 0.17 & 0.16 &
0.16 \\ \hline
$g$ & 0.37 & 1.82 & -4.81 & -2.40 & 0.05 & 0.05 & 0.50 & 0.25 & 0.07 &
0.07 \\ \hline
\end{tabular}
\caption{Coefficients of the extrapolation of the input fragmentation
functions at small $x$. Column 2 gives the cut--off $x_0$ where we
switch from the FFs from \cite{Poetter} to a parameterization in the
form (\ref{fpara}). Columns 3 to 5 give the coefficients of this
parameterization, as applied to the sum $\sum_h D_i^h$. The remaining
columns give normalizations $N_i^h$, so that $D_i^h = N_i^h \, \sum_h
D_i^h$. Note that $h$ always stands for the sum of particle and
anti--particle, whenever the two are not identical; for example,
$\pi^\pm$ stands for the sum of $\pi^+$ and $\pi^-$, $K^0$ stands for
the sum of $K^0$ and $\overline{K^0}$, etc. }
\label{extrapolation}
\end{center}
\end{table}
\end{center}

\setcounter{equation}{0}
\renewcommand{\theequation}{E.\arabic{equation}}

\section*{Appendix E: Stable particle spectra for different initial
(super)particles} 
\label{sec:Curves}

Here we give an almost complete set of FFs for different initial
particles, for one set of SUSY parameters, with low $\tan\beta$ and
gaugino--like LSP; the dependence of these results on the SUSY
parameters has been analyzed in Sec.~\ref{subsec:SUSY_dep}. We
used a ratio of Higgs vevs $\tan \beta = 3.6$, a gluino and scalar
mass scale $M_{\rm SUSY} \sim 500$ GeV, a supersymmetric Higgs mass
parameter $\mu = 500$ GeV, a CP--odd Higgs boson mass $m_A = 500$ GeV
and trilinear soft breaking parameter $A_t = 1$ TeV. As usual, we
plot $x^3 \cdot D_I^P (x,M_X)$. We take $M_X = 10^{16}$ GeV, as
appropriate for a GUT interpretation of the $X$ particle.

\clearpage


\begin{figure}
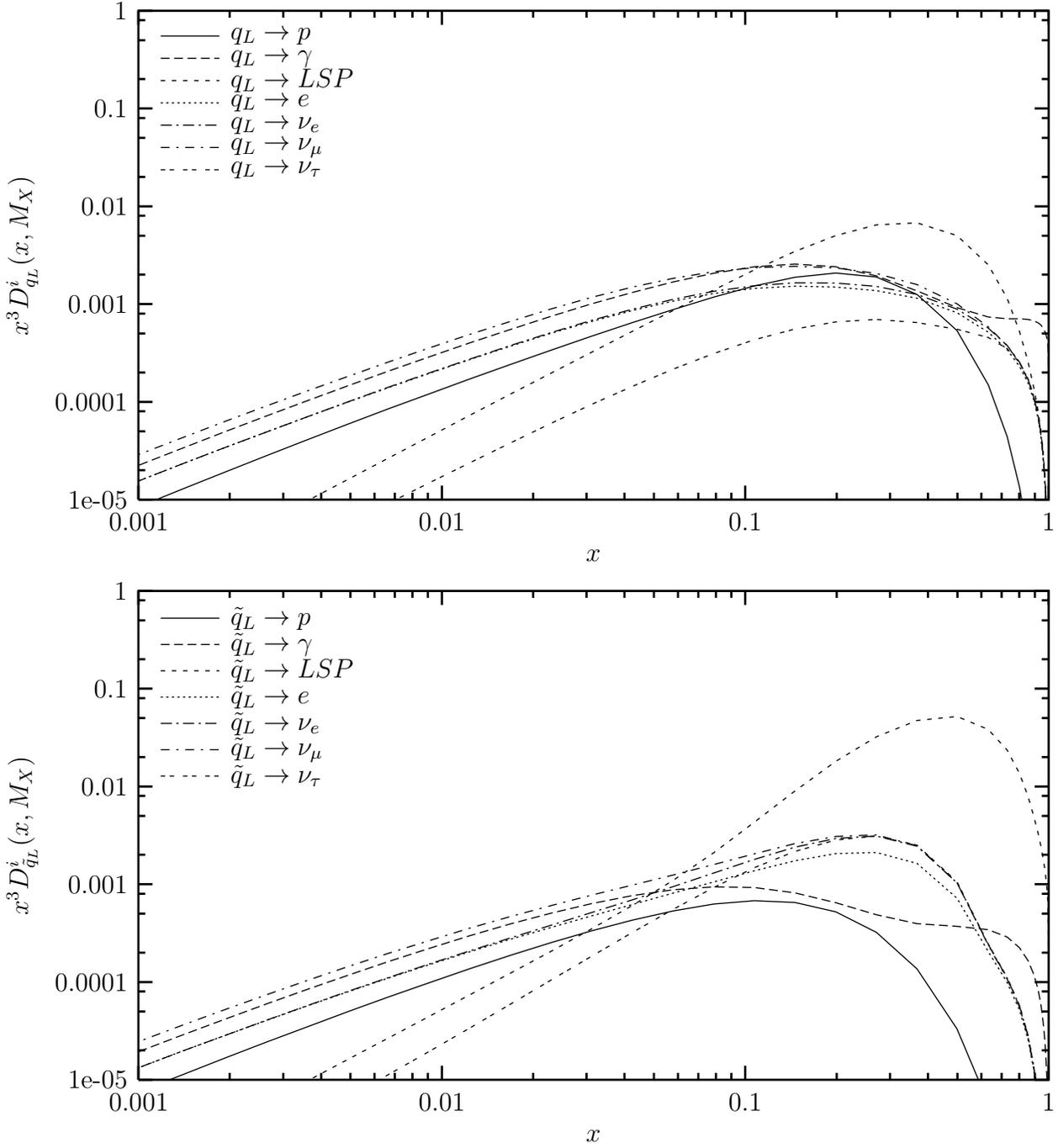

\input{Low_G_uL.tex}
\input{Low_G_uL_.tex}
\caption{Fragmentation functions of a first or second generation
  $SU(2)$ doublet quark (top) and a squark (bottom) into stable
  particles.}
\end{figure}

\begin{figure}
\input{Low_G_uR.tex}
\input{Low_G_uR_.tex}
\caption{Fragmentation functions of a first or second generation
  $SU(2)$ singlet quark (top) and a squark (bottom) into stable
  particles.}
\end{figure}

\begin{figure}
  \input{Low_G_tL.tex} \input{Low_G_tL_.tex}
\caption{Fragmentation functions of a third generation $SU(2)$ doublet
quark (top) and a squark (bottom) into stable particles.} 
\end{figure}

\begin{figure}
\input{Low_G_tR.tex}
\input{Low_G_tR_.tex}
\caption{Fragmentation functions of a third generation $SU(2)$ singlet
quark (top) and a squark (bottom) into stable particles.} 
\end{figure}

\clearpage


\begin{figure}
\input{Low_G_eL.tex}
\input{Low_G_eL_.tex}
\caption{Fragmentation functions of a first or second generation
$SU(2)$ doublet lepton (top) or slepton (bottom) into stable
particles. The structures in some of the curves in the lower frame
originate from 2--body decay kinematics.}
\end{figure}

\begin{figure}
\input{Low_G_eR.tex}
\input{Low_G_eR_.tex}
\caption{Fragmentation functions of a first or second generation
$SU(2)$ singlet lepton (top) or slepton (bottom) into stable
particles. The structures in some of the curves in the lower frame
originate from 2--body decay kinematics.}
\end{figure}

\begin{figure}
\input{Low_G_tauL.tex}
\input{Low_G_tauL_.tex}
\caption{Fragmentation functions of a third generation
$SU(2)$ doublet lepton (top) or slepton (bottom) into stable
particles. The structures in some of the curves in the lower frame
originate from 2--body decay kinematics.}
\end{figure}

\begin{figure}
\input{Low_G_tauR.tex}
\input{Low_G_tauR_.tex}
\caption{Fragmentation functions of a third generation
$SU(2)$ singlet lepton (top) or slepton (bottom) into stable
particles. The structures in some of the curves in the lower frame
originate from 2--body decay kinematics.}
\end{figure}

\clearpage


\begin{figure}
\input{Low_G_B.tex}
\input{Low_G_B_.tex}
\caption{Fragmentation functions of a $B$ boson(top) and a Bino
(bottom) into stable particles.}
\end{figure}

\begin{figure}
\input{Low_G_W.tex}
\input{Low_G_W_.tex}
\caption{Fragmentation functions of a $W$ boson (top) and a Wino
(bottom) into stable particles.}
\end{figure}

\begin{figure}
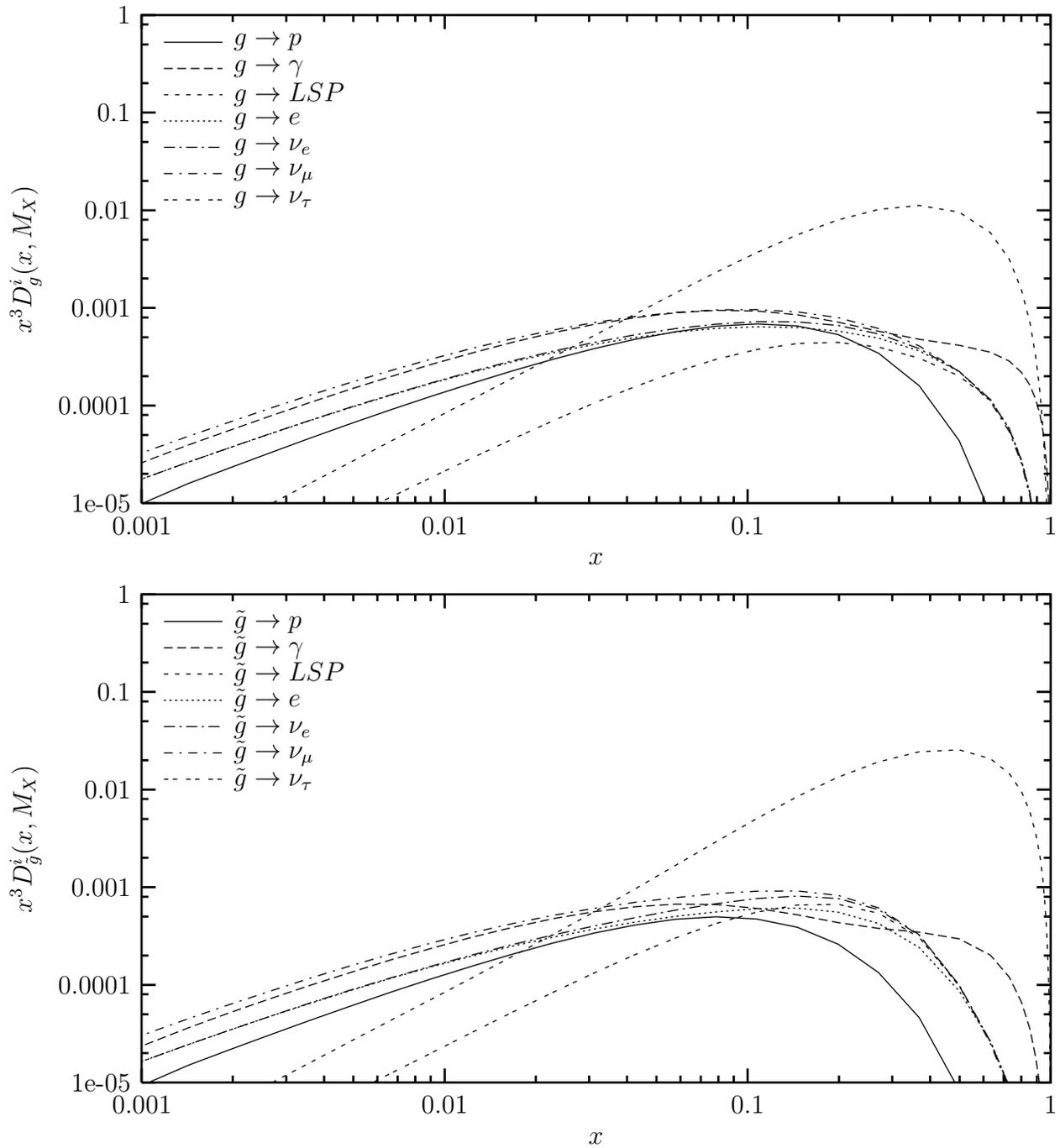

\input{Low_G_g.tex}
\input{Low_G_g_.tex}
\caption{Fragmentation functions of a gluon (top) and a gluino
(bottom) into stable particles.}
\end{figure}

\clearpage

\begin{figure}
\input{Low_G_H1.tex}
\input{Low_G_H1_.tex}
\caption{Fragmentation functions of a $H_1$ Higgs boson (top) and a
$\tilde{H}_1$ higgsino (bottom) into stable particles.}
\end{figure}

\begin{figure}
\input{Low_G_H2.tex}
\input{Low_G_H2_.tex}
\caption{Fragmentation functions of a $H_2$ Higgs boson (top) and a
$\tilde{H}_2$ higgsino (bottom) into stable particles.}
\end{figure}

\clearpage


\end{document}